\documentclass[11pt]{article}

\usepackage{epsfig}
\usepackage{amsfonts,amssymb}

\newcommand{\sect}[1]{\setcounter{equation}{0}\section{#1}}
\renewcommand{\theequation}{\arabic{section}.\arabic{equation}}

\newcommand{\al}{\ensuremath{\alpha}}
\newcommand{\ga}{\ensuremath{\gamma}}
\newcommand{\Ga}{\ensuremath{\Gamma}}

\newcommand{\ep}{\ensuremath{\epsilon}}
\newcommand{\vep}{\ensuremath{\varepsilon}}

\newcommand{\la}{\ensuremath{\lambda}}
\newcommand{\La}{\ensuremath{\Lambda}}
\newcommand{\om}{\ensuremath{\omega}}
\newcommand{\Om}{\ensuremath{\Omega}}
\newcommand{\p}{\ensuremath{\phi}}
\newcommand{\s}{\ensuremath{\sigma}}

\renewcommand{\t}{\ensuremath{\tau}}
\renewcommand{\th}{\ensuremath{\theta}}

\newcommand{\D}{\ensuremath{{\cal D}}}

\renewcommand{\L}{\ensuremath{{\cal L}}}
\newcommand{\M}{\ensuremath{{\cal M}}}
\newcommand{\N}{\ensuremath{{\cal N}}}
\renewcommand{\O}{\ensuremath{{\cal O}}}

\renewcommand{\d}{\ensuremath{{\rm d}}}

\newcommand{\ra}{\ensuremath{\rightarrow}}
\newcommand{\del}{\ensuremath{\partial}}

\newcommand{\td} {\ensuremath{\tilde}}
\newcommand{\inv}{\ensuremath{^{-1}}}

\newcommand{\Tr}{\ensuremath{{\rm Tr}}}
\newcommand{\pp}{\ensuremath{{p^{+}}}}

\newcommand{\sxi}{\ensuremath{\not\!\xi}}

\newcommand{\be}{\begin{equation}}
\newcommand{\ee}{\end{equation}}

\newcommand{\ba}{\begin{eqnarray}}
\newcommand{\ea}{\end{eqnarray}}

\newpage

\begin{document}

\newpage

\bigskip
\hskip 4.8in\vbox{\baselineskip12pt
\hbox{hep-th/0206045}}

\bigskip
\bigskip
\bigskip

\begin{center}
{\Large \bf Penrose Limits, Deformed pp--Waves}\\
\bigskip
{\Large \bf and the}\\
\bigskip
{\Large \bf String Duals of ${\cal N}=1$ Large $N$ Gauge Theory}

\end{center}
\bigskip
\bigskip
\bigskip

\centerline{\bf Dominic Brecher$^\natural$, Clifford
 V. Johnson$^\natural$, Kenneth J. Lovis$^\natural$, Robert
 C. Myers$^\sharp$}

\bigskip
\bigskip
\bigskip

\centerline{\it $^\natural$Centre for
Particle Theory, Department of Mathematical Sciences}
\centerline{\it University of Durham, Durham, DH1 3LE, U.K.}
\centerline{\small \tt dominic.brecher, c.v.johnson,
  k.j.lovis@durham.ac.uk} \centerline{$\phantom{and}$}
\centerline{\it$^\sharp$Perimeter Institute for Fundamental Physics,
  35 King Street North} \centerline{\it Waterloo, Ontario N2J 2VW,
  Canada} \centerline{\it Department of Physics, University of
  Waterloo,} \centerline{\it Waterloo, Ontario N2L 3G1, Canada}
\centerline{\it Physics Department, McGill University, 9600 University
  Street} \centerline{\it Montr\'eal, Qu\'ebec H3A 2T8, Canada}


\centerline{\small \tt  rcm@hep.physics.mcgill.ca}

\bigskip
\bigskip


\begin{abstract}
  \vskip 2pt A certain conformally invariant ${\cal N}=1$
  supersymmetric $SU(N)$ gauge theory has a description as an
  infra--red fixed point obtained by deforming the ${\cal N}=4$
  supersymmetric Yang--Mills theory by giving a mass to one of its
  ${\cal N}=1$ chiral multiplets.  We study the Penrose limit of the
  supergravity dual of the large~$N$ limit of this ${\cal N}=1$ gauge
  theory.  The limit gives a pp--wave with R--R five--form flux
  and both R--R and NS--NS three--form flux. We discover that this new
  solution preserves twenty supercharges and that, in the light--cone
  gauge, string theory on this background is exactly solvable.
  Correspondingly, this latter is the stringy dual of a particular
  large charge limit of the large $N$  gauge theory.  We are able
  to identify which operators in the field theory survive the limit to
  form the string's ground state and some of the spacetime
  excitations. The full string model, which we exhibit, contains a
  family of non--trivial predictions for the properties of the gauge
  theory operators which survive the limit.
\end{abstract}
\newpage
\baselineskip=18pt
\setcounter{footnote}{0}


\sect{Introduction and Conclusions}

The AdS/CFT correspondence~\cite{adscft1,adscft2,adscft3} has given us
a concrete example of how to realise the old expectation~\cite{thooft}
that large~$N$ $SU(N)$ gauge theory may be written as a theory of
strings. Unfortunately, most of the direct computations on the
Anti--de Sitter (AdS) side of the correspondence are not inherently
stringy.  This is largely due to the fact that the string theory
background contains a non--trivial Ramond--Ramond (R--R) five--form
flux, and the quantisation of string theory in such backgrounds is a
task not yet fully understood in generality. Instead, much use has
been made of the supergravity limit for reliable computations.

Nevertheless, there are many computations which have provided useful
links between the supergravity truncation and the full superstring
theory. Examples are those involving and, indeed, {\it requiring}
explicit recourse to the extended nature of the branes which underlie
the geometry. Branes expanding {\it via} the dielectric
mechanism~\cite{gorob} into giant gravitons~\cite{giants} are needed
to understand the ``stringy exclusion principle''~\cite{exclude}.  It
has been partially shown~\cite{joematt} how the same mechanism
underlies the process whereby the type IIB theory produces the rich
family of vacua corresponding to the $\N = 4$ Yang--Mills theory broken
to ${\cal N}=1$ by adding masses to all three chiral
multiplets~\cite{VW}.  Meanwhile the enhan\c con mechanism~\cite{jpp}
has been shown~\cite{bpp,ejp} to be crucial in supplementing singular
supergravity duals~\cite{FGPW,pw} in order to yield the correct description
of the ${\cal N}=2$ gauge theory vacua.  Finally, while supergravity techniques are sufficient to calculate the
wavefunctions associated with various glueball states --- see, for
example, refs.~\cite{glu1,glu2} ---
the soft high energy behaviour of string scattering amplitudes
is an essential ingredient in matching expected behavior for
scattering these states in the dual gauge theory~\cite{joematt2}.

While these examples and many others in the same spirit are
interesting scenarios in which to observe the truly stringy nature of
the AdS/CFT correspondence at work, they are still indirect, and help
only in characterising the available vacua of the theory.  They
provide only a small window of understanding of many dynamical issues;
for this, one would need to consider aspects of the full string theory
in the AdS background.

Important progress has recently been made however, at least in the
case of a much simpler background.  It has been shown~\cite{BFHP1}
that the type IIB supergravity theory has another maximally
supersymmetric solution besides Minkowski space and AdS$_5\times S^5$.
This solution is a ``pp--wave'' with a null R--R five--form flux
switched on. As anticipated in ref.~\cite{BFHP1}, it seems natural that
such a highly symmetric solution should have a role in the AdS/CFT
discussion.  Furthermore, it was shown~\cite{metsaev} that string
theory in this background is exactly solvable in the light--cone
gauge. In fact, it was discovered~\cite{BFHP2,BMN,BFP} that there is a
direct connection to be made: pp--waves can be obtained from any
supergravity solution \emph{via} the so--called ``Penrose
limit''~\cite{penrose,guven}, and in fact the maximally supersymmetric
wave is just such a limit of the AdS$_5\times S^5$ geometry.

Correspondingly, one expects that there is a limit of the dual gauge
theory which can be identified as the dual of type IIB string theory
on the resulting maximally supersymmetric pp--wave. Since the full
type IIB string theory in this background is solvable, one might also
hope to finally be able to make direct and highly stringy statements
about the gauge theory. These expectations have been borne out by the
direct identification~\cite{BMN} of the set of operators which survive
the limit, and furnish the string ground state and the spectrum of
excitations. The relevant limit is to consider the sector charged
under a $U(1)_R$ subgroup of the full $SO(6)_R$ R--symmetry of the
theory, and take the charge $J$ to be extremely large, growing as
$N^{1/2}$.  Sensible gauge theory results are obtained if at the same
time $g_{\rm YM}^2$ is kept small and fixed and hence the 't Hooft
coupling $\lambda=g^2_{\rm YM}N$ is sent to infinity. Since the 't
Hooft coupling corresponds to the curvature scale in the string dual,
which is $L=(g^2_{\rm YM}N)^{1/4}$ in units of the string coupling,
this gauge theory limit fits nicely with the Penrose limit on the
geometry, which also (as we will review) takes $L\to\infty$. We should
now compare stringy results in the pp--wave background with gauge
theory results in this limit.  For example, the structure of the
string Hamiltonian alone makes a highly non--trivial prediction for
the behaviour of the anomalous dimensions of a particular set of non--
(but near--) BPS operators in the theory at large 't Hooft coupling:
\begin{eqnarray}
(\Delta-J)_n= \sqrt{1+\frac{n^2 g_{\rm YM}^2N}{J^2}}=1+\frac{n^2
g_{\rm YM}^2N}{2 J^2}+\cdots
\label{anomalous}
\end{eqnarray}
Here, $\Delta$ is the dimension of the operator and $n$ labels an
excitation level in the string theory on the one hand and, on the other, a
particular type of operator in the dual gauge theory made by
constructing ordered ``words'' of strings of field insertions under
the gauge trace (see ref.~\cite{BMN}). The above prediction has been checked
to leading order~\cite{BMN} in an appropriate new expansion
parameter,~$\lambda^\prime=g_{\rm YM}^2N/J^2$.

The hope is that the example above (and others in that spirit) will
teach us new lessons about the AdS/CFT correspondence and about other
more general gravity/gauge theory correspondences. The aim of this
paper is to understand the meaning of this new facet of the
correspondence in the context of a particularly interesting and useful
class of backgrounds. These are the backgrounds which
represent~\cite{Girardello:1998pd,Distler:1998gb} the ``Holographic
Renormalisation Group (RG) Flow'' from the $\N = 4$ Yang--Mills theory
to other gauge theories of interest, by a controlled deformation of the
AdS$_5\times S^5$ background. We have used what is considered (by
some) as one of the cleanest and most instructive examples of such
geometries, the flow to an infra--red (IR) $\N=1$ supersymmetric fixed
point, in which part of the dual's geometry is again conformal to
AdS$_5$. While we have obtained some interesting results, and surmount
a number of technical obstructions, we do not yet have a satisfactory
understanding of all of our results, for reasons which will become
clear later.

So for the main part of our presentation, we shall be concerned
with a study of the endpoint of the flow, showing that the Penrose
limit of the supergravity solution is again (not surprisingly) a
pp--wave, but it is significantly different from previous examples
which have been related to known gauge theories: in addition to the
R--R five--form flux, it has R--R and Neveu--Schwarz--Neveu--Schwarz
(NS--NS) three--form flux. It is therefore not maximally
supersymmetric as in other recent examples~\cite{IKM,GO,PS} dual to
large charge ${\cal N}=1$ large $N$ gauge theories, but instead
preserves twenty supercharges, possessing some of the unusual number
of extra or ``supernumerary'' supersymmetries~\cite{CLP1,CLP2,GH}
beyond the standard half that pp--waves have generically.

Such a strange number of supersymmetries, as we predict here for a
four--dimensional gauge theory is not immediately implausible. Recall
that we have two special features: four--dimensional conformal
invariance, and a special large charge limit which treats a particular
direction in R--symmetry space as special.  One achieves 32
supercharges in the usual context by first having the na\"{\i}vely maximal
sixteen, and then observing that the closure with the conformal algebra
doubles each of the four supersymmetries (each equivalent to a single
$\N=1$ supersymmetry in four dimensions). It is
conceivable that since we have picked a special direction out in
R--symmetry space, there is a scaling limit on the superalgebra which
can ensure that only one of the four supercharges gets
doubled by demanding closure with the conformal algebra, giving the twenty
we find here.\footnote{This is so far a conjecture as to how our
  prediction from the Penrose limit is realised, and it probably
  amounts to an Inonu--Wigner contraction of the full four--dimensional
  superconformal algebra. It is comforting to note that such limits
  are known to yield interesting subalgebras of the superconformal
  group in two dimensions, (see for example refs.~\cite{Ali:sd} and
  references therein), and so it is worth exploring in four dimensions.
  For a recent discussion which may be relevant, see
  ref.~\cite{Arutyunov:2002xd}. }

Another direct prediction of our new model comes from deriving the
Hamiltonian of the string theory and comparing it to gauge theory in
an analogous manner to the formula in equation~(\ref{anomalous}).  While
we have a sector of the operator spectrum which produces a formula
like the above (with an appropriate redefinition of $J$ to what we
shall call $\hat{J}$; see later), we have a much more intriguing
formula from another sector of the theory:
\be ( \Delta - \hat{J}
)_n^{\pm} = \sqrt{ \frac{5}{2} + \frac{n^2 g_{YM}^2 N}{\hat{J}^2}
  \pm \frac{1}{2} \sqrt{9 + 12 \frac{n^2 g_{YM}^2 N}{\hat{J}^2}
    }}.
\ee
Not only is the double square root an amusing challenge for the gauge
theory side, observe further the form of the leading terms in the expansion:
\begin{eqnarray}
( \Delta - \hat{J}
)_n^{+}&=&2+\frac{n^2 g_{\rm YM}^2N}{2\hat{J}^2}+\cdots\nonumber\\
( \Delta - \hat{J}
)_n^{-}&=&1+\frac{1}{6}\left(\frac{n^2 g_{\rm YM}^2N}{\hat{J}^2}\right)^2+\cdots
\end{eqnarray}
We see that the latter sector contains a contribution which appears at
second order in a $\lambda^\prime$ expansion, there being an exact
cancellation at linear order. This either corresponds to a new class
of diagrams on the gauge theory side, or a new exact cancellation in
the usual diagrams.  Since our exact string theory model is predicted
to be dual to a known gauge theory, it will be a very interesting task
to verify this behaviour.

The paper is organised as follows.  We start our detailed discussion
in section \ref{sec:adding}, where we consider the addition of a
three--form to the supersymmetric pp--waves of type IIB
supergravity~\cite{BFHP1}.  The resulting solutions cannot be
maximally supersymmetric, but they are always at least one--half
supersymmetric.  Moreover, solutions preserving other exotic fractions
can be constructed with relative ease.  Although we do not undertake
an exhaustive classification of all possibilities here, we present the
specific solution which preserves 20 of the 32 supersymmetries; this
seems to be the largest number that can be preserved in this case.

In section \ref{sec:IR}, we consider a Penrose limit of the
ten--dimensional fixed point geometry presented by Pilch and
Warner~\cite{PW1}, focusing on a null geodesic with large angular
momentum in one of the directions corresponding to part of the moduli
space of the gauge theory.  The resulting solution includes a
non--trivial three--form, and we find precisely the 5/8
supersymmetric pp--wave of section~\ref{sec:adding}.

Our next step is to analyse the string spectrum in this background,
which we do in section~\ref{sec:spectrum}.  String theory in this
background is considerably more complicated than in the maximally
supersymmetric case, studied in refs.~\cite{metsaev,metsaevtseytlin}.
Moreover, although the effect of the three--form on the string
spectrum is similar to cases considered in refs.~\cite{forgacs,BMN,RT},
there are important differences.  Since it is very interesting to have
(potentially) the complete string theory dual of a sector of a
four--dimensional gauge theory away from maximal supersymmetry, we
remark on a number of features which may have some significance for
the dual gauge theory.

In section \ref{sec:ft}, we turn our attention to the study of the
lowest lying sectors of the string theory and their comparison to a
particular large charge limit of the ${\cal N}=1$ superconformal gauge
theory, which we of course conjecture to be dual to type IIB string theory
in this background.

While we have not fully understood the physical interpretation (in
terms of a dual gauge theory) of our results for the entire flow
geometry~\cite{PW2} representing the RG flow from the $\N = 4$
Yang--Mills theory to the IR fixed point, we present its Penrose limit
in section~\ref{sec:flow}. The resulting pp--wave is one--half
supersymmetric at any point along the flow.  Taking the Penrose limit
thus enhances supersymmetry from 1/8 to 1/2 at any point along the
flow geometry, and from 1/4 to 5/8 at the IR fixed point.

One of the technicalities to be faced in this RG flow context is the
fact that such geometries are written naturally in Poincar\'e
coordinates, while the Penrose limits of recent interest are mainly
performed on AdS spaces written in terms of global coordinates. This
leads to the interpretational difficulties alluded to above, since a
simple trajectory in global coordinates representing a state (or class
of states) in the field theory translates into a trajectory which
probes a wide range of energy scales in the field theory. It is
puzzling to interpret this cleanly in the dual field theory.

For completeness, and to summarise some of our conventions, we give an
exposition of the ten--dimensional Pilch--Warner
geometry~\cite{PW1,PW2} in appendix A.  In appendix B, we present a
brief discussion of an alternate set of geodesics ($\theta=\pi/2$)
and the associated Penrose limit of the IR geometry.  Appendix C discusses
a potential instability of superstring theory in the pp--wave
backgrounds presented in section \ref{sec:adding_soln}, and finally appendix D reviews
the Penrose limit of AdS$_5\times S^5$ in Poincar\'{e}
coordinates~\cite{BFP,KP}.

\bigskip\bigskip

{\bf Note added in preparation:} The results reported here were
presented by C.V.J. at the workshop on Strings and Branes, held at
KIAS, Korea, on May 29th of this year. While subsequently preparing
this manuscript for publication, refs.~\cite{CHKW} and \cite{gimon}
appeared, which have some overlap with our results.


\sect{Adding three-forms to the supersymmetric IIB pp--wave}
\label{sec:adding}

\subsection{The generic solution}
\label{sec:adding_soln}

Since we want to retain the possibility of being able to exactly
quantise superstring theory propagating on our pp--waves, we will only
consider backgrounds with constant dilaton and axion.  Without loss of
generality we can take both to vanish.  In that case, the bosonic
sector of type IIB supergravity contains the metric, an R--R
four--form potential, $C_4$, with self--dual field strength, and NS--NS
and R--R two--form potentials, $B_2$ and $C_2$, respectively.
Combining these latter as $A_2 = B_2 + i C_2$, with field strength
$G_3 = \d A_2 = H_3 + i F_3$, the self--dual five--form is given by\footnote{
We take the field equations (\ref{eqn:iib}) and supersymmetry
transformations (\ref{eqn:delta_fermions}) of type IIB
supergravity from ref.~\cite{schwarz:83}.  Our conventions are different,
however, in that we use a ``mostly plus'' signature, but still have that
$\{\Ga_a,\Ga_b\}=2g_{ab}$.  To write the
results of ref.~\cite{schwarz:83} in our conventions, we thus send $g_{ab} \ra
-g_{ab}$ and $\Ga_a \ra i \Ga_a$, so that all our gamma matrices are
real.  }
\be
F_5 = \star F_5 = \d C_4 - \frac{1}{8} \mbox{Im} \left( A_2 \wedge G_3^*
\right),
\ee

\noindent where $*$ denotes complex conjugation and $\star$ denotes
ten--dimensional Hodge duality.  The equations
of motion are then~\cite{schwarz:83}
\[
R_{ab} = \frac{1}{6} F_{acdef} F_b^{~cdef} + \frac{1}{8} \left(
G_{acd} G_b^{*cd} + G_{acd}^* G_b^{~cd} -\frac{1}{6} g_{ab} G_{cde}
G^{*cde} \right),
\]
\be
\d \star G_3 = 4i F_5 \wedge G_3, \qquad G_{abc} G^{abc} = 0,
\label{eqn:iib}
\ee
\[
\d \star F_5 = \d F_5 = -\frac{1}{8} \mbox{Im} \left( G_3 \wedge G_3^*
\right),
\]

\noindent and the remaining Bianchi identity is $\d G_3 = 0$.  The
fermionic sector of the theory consists of a dilatino, $\la$, and
gravitino, $\psi_a$, with opposite chirality.  Taking $\Ga_{11} = \Ga_0
\Ga_1 \ldots \Ga_9$, we have
\be
\Ga_{11} \la = -\la, \qquad \Ga_{11} \psi_a = \psi_a.
\ee

\noindent The supersymmetry transformations are given in terms
of a single complex chiral spinor, $\vep$, of the same positive
chirality as the gravitino, and which can be written in terms of two
sixteen-component Majorana--Weyl spinors: $\vep=\theta^1+i\theta^2$.  For purely
bosonic backgrounds one has~\cite{schwarz:83}
\ba
\delta \la &=& \frac{1}{24} G_{abc} \Ga^{abc} \vep, \nonumber \\
\delta \psi_a &=& D_a \vep - \Om_a \vep - \La_a \vep^*,
\label{eqn:delta_fermions}
\ea

\noindent where we have defined
\be
\Om_a = -\frac{i}{480} F_{b_1 \ldots b_5} \Ga^{b_1 \ldots b_5} \Ga_a,
\qquad \La_a =  \frac{1}{96} \left( G_{bcd} \Ga_a^{~bcd} - 9 G_{abc}
\Ga^{bc} \right),
\label{eqn:susy}
\ee

\noindent for later use.

The one--half supersymmetric pp--wave solution of the type IIB
theory presented in ref.~\cite{BFHP1} is
\ba
\d s^2 &=& 2 \d u \d v + H(u,x) \d u^2 + \d s^2(\mathbb{E}^8),
\label{eqn:iibwave} \\
F_5 &=& (1 + \star) \d u \wedge \om_4, \nonumber
\ea

\noindent where for each $u$, the four--form $\om_4 (u,x)$ is closed and
co--closed (it can depend on $u$ in an arbitrary way).  The equations
of motion relate $H$ and $\om_4$ as
\be
\nabla^2 H = -\frac{2}{3} \om_4^2 \equiv -\frac{2}{3} \om_{ijkl} \om^{ijkl},
\ee

\noindent where $x^i, i=1,\ldots,8,$ are the coordinates, and
$\nabla^2$ the Laplacian, on $\mathbb{E}^8$.  Taking
$H=A_{ij} x^i x^j$, with $A_{ij}$ a constant
symmetric matrix, gives a Cahen--Wallach space~\cite{CW}.
With $\om_4 = \mu \ep (\mathbb{E}^4)$ a constant multiple of the volume form
on one of the transverse $\mathbb{E}^4$s, there is
a unique choice\footnote{Up to the differences in conventions between
our work and that of ref.~\cite{BFHP1}.} of $A_{ij} = -\mu^2 \delta_{ij}$ for which the solution is maximally
supersymmetric~\cite{BFHP1}.  Switching on further constant components
of $\om_4$ gives rise to pp--waves preserving between 16 and 32
supersymmetries~\cite{CLP1,CLP2,GH}, and some of these more general pp--waves further
arise as Penrose limits of various intersecting brane
solutions~\cite{BFP,CLP2,LV}.

It is straightforward to switch on a non--trivial three--form, but one will
then lose \emph{maximal} supersymmetry.  Since the Ricci
scalar for the metric (\ref{eqn:iibwave}) vanishes, we have
$G_{abc}G^{*abc}=0$.  Combining this with the equation of motion $G_{abc}G^{abc}=0$, we must
have that both $F_3$ and $H_3$ are independently null, in which
case the three--form can be written as
\be
G_3 = \d u \wedge \xi_2,
\label{eqn:wave_3-form}
\ee

\noindent for some complex two--form $\xi_2=\alpha_2(u,x) + i
\beta_2(u,x)$.  Then, since $F_5 \wedge G_3 = G_3 \wedge G_3^* = 0$,
all of the the equations (\ref{eqn:iib}) will be
satisfied if, for each $u$, both $\om_4$ and $\xi_2$ are closed and co--closed, and if
\be
\nabla^2 H = -\frac{2}{3} \om_4^2 - \frac{1}{2} \left|
\xi_2 \right|^2,
\label{eqn:box_H}
\ee

\noindent where $\left| \xi_2 \right|^2 = \xi_{ij} \xi^{*ij}$.

One again finds that this generic solution preserves one--half of the
32 supersymmetries, those which satisfy $\Ga^u \vep = 0$ in the usual
fashion.  Other exotic fractions are possible, however, and
although we leave a systematic study for future work, we will show
here how to construct a solution which preserves 20
supersymmetries.


\subsection{A 5/8 supersymmetric wave}
\label{sec:adding_susy}

We now specialize to the following:
\ba
\d s^2 &=& 2 \d u \d v + A_{ij} x^i x^j \d u^2 + \d s^2(\mathbb{E}^8), \nonumber \\
F_5 &=& \mu ~(1 + \star) \d u \wedge \ep (\mathbb{E}^4),
\label{eqn:ansatz} \\
G_3 &=& \zeta ~\d u \wedge \d z^1 \wedge \d z^2, \nonumber
\ea

\noindent where $z^1=x^1+ix^2, z^2=x^3+ix^4$ are complex coordinates
on one of the transverse $\mathbb{E}^4$s, $\mu$ and $\zeta$ are real and complex constants
respectively, and the equation of motion (\ref{eqn:box_H}) demands that
\be
{\rm tr} A = -8 \mu^2 -2 |\zeta|^2.
\label{eqn:trace_A}
\ee

\noindent As usual, we make use of the off--diagonal
orthonormal basis
\be
e^- = \d u, \qquad e^+ = \d v + \frac{1}{2} A_{ij} x^ix^j \d u, \qquad
e^i = \d x^i,
\ee

\noindent with metric $\eta_{+-} = 1$ and $\eta_{ij} = \delta_{ij}$.
The only non-trivial component of the spin connection is then
\be
\om_{-i} = \om^{+i} = A_{ij} x^j \d u.
\ee

Consider, first, the dilatino variation
\be
\delta \la = \frac{1}{8} \sxi~ \Ga_+ \vep = 0,
\label{eqn:dilatino}
\ee

\noindent where we have defined $\sxi = \xi_{ij} \Ga_{ij}$.  Any
Killing spinors must thus satisfy
\be
\left( 1-\Ga_0\Ga_9 \right) \left( 1 + i\Ga_1 \Ga_2 \right) \left(1 +
i\Ga_3\Ga_4 \right) \vep = 0,
\ee

\noindent where we take the pp--wave to be moving in the $x^9$
direction.  This equation has 28 independent solutions,
characterised by the eigenvalues of the boost operator,
$\Ga_0\Ga_9$, and rotation operators, $i\Ga_1\Ga_2$ and
$i\Ga_3\Ga_4$.  Schematically, the solutions are
\ba
(i)&&(+,\pm ,\pm ,\pm ,\pm ),
\nonumber\\
(ii)&&(-,-,+,\pm ,\pm ),
\nonumber\\
(iii)&&(-,+,-,\pm,\pm ),
\label{eqn:dil_soln}\\
(iv)&&(-,-,-,\pm ,\pm ),
\nonumber
\ea

\noindent where the first three entries denote the eigenvalues ($\pm1$) of the
operators $\Ga_0\Ga_9$, $i\Ga_1\Ga_2$ and $i\Ga_3\Ga_4$ respectively.
The eigenvalues of the remaining
two rotation operators $i\Ga_5\Ga_6$ and $i\Ga_7\Ga_8$ are, of course,
arbitrary.\footnote{Of course, the overall product of eigenvalues must be positive
as implied by the chirality projection $\Gamma_{11}\varepsilon=\varepsilon$.}
  Our pp--wave will thus preserve at least 16, but at most 28,
supersymmetries.  There are a possible
12 supernumerary Killing spinors (cases ($ii$--$iv$)), annihilated by $\Ga_-$, in addition to
the usual 16 (case ($i$)), annihilated by $\Ga_+$.  Moreover, it would seem that this is the largest
number of supersymmetries that will be preserved by such solutions,
since switching on components of $G_3$ in additional directions will only
lead to fewer solutions of (\ref{eqn:dilatino}).

Further conditions on the Killing spinors come from the variation of
the gravitino.  Using the analysis of ref.~\cite{FP,BFHP1} as a guide, we have
\ba
D_a \vep &=& \Om_a \vep + \La_a \vep^*, \nonumber \\
\Om_a &=& -\frac{i\mu}{4} \left( \Ga_{1234} + \Ga_{5678}
\right) \Ga_+ \Ga_a, \label{eqn:gravitino} \\
\La_a &=& \frac{1}{32} \left( \xi_{ij} \Ga_{a+ij} - 3
G_{abc} \Ga^{bc} \right). \nonumber
\ea

\noindent Since $\Om_v = \La_v = 0$, the $v$ component implies $\vep =
\vep (u,x)$, and the $i$ components give
\ba
\del_i \vep &=& \Om_i \vep + \La_i \vep^*, \nonumber \\
\Om_i &=& -\frac{i \mu}{4} \left( \Ga_{1234} + \Ga_{5678}
\right) \Ga_+ \Ga_i, \\
\La_i &=& \frac{1}{32} \left( \Ga_i \sxi - 8
\xi_{ij} \Ga_j \right) \Ga_+. \nonumber
\ea

\noindent Using the fact that $\Ga_+^2=0$, we have
\be
\Om_i \Om_j = \La_i \La_j = \Om_i \La_j = 0
\label{eqn:squares}
\ee

\noindent so that
\be
\vep = \chi(u) + x^i \rho_i (u).
\ee

\noindent Differentiating with respect to $x^i$, one finds that
$\rho_i = \Om_i \chi + \La_i \chi^*$, so
\be
\vep = \chi + x^i ( \Om_i \chi + \La_i \chi^* ).
\label{eqn:formm}
\ee

\noindent Note that the condition (\ref{eqn:dilatino}) now acts only on
$\chi$: $\sxi ~\Ga_+ \chi = 0$.

We are left with the $u$ component of (\ref{eqn:gravitino}):
\ba
\del_u \vep &=& \frac{1}{2} A_{ij} x^i \Ga_j \Ga_+ \vep + \Om_u \vep +
\La_u \vep^*, \nonumber \\
\Om_u &=& -\frac{i}{4}\mu \left( \Ga_{1234} + \Ga_{5678} \right) \Ga_+
\Ga_-, \label{eqn:lastt}\\
\La_u &=& -\frac{1}{32} \sxi \left( \Ga_+ \Ga_- + 2
\right). \nonumber
\ea

\noindent Given the form (\ref{eqn:formm}) of the spinors, the constraint
(\ref{eqn:lastt}) may be divided into separate components by
collecting the terms independent of $x^i$
and those linear in each of the $x^i$. Making use of (\ref{eqn:squares}), the
former may be written as:
\be
\frac{ \d \chi}{\d u} = -\frac{i}{2}\mu \left( \Ga_{1234} + \Ga_{5678} \right) \chi -
\frac{1}{32} \sxi~ \left( \Ga_+ \Ga_- + 2 \right) \chi^*.
\ee
Using this result, the remaining linear terms can be reduced to
eight algebraic constraints
\[
\left( A_{ij} \Ga_j \Ga_+ + \mu^2 \Ga_i \Ga_+ - \frac{1}{4}
\sxi~ \La^*_i + \frac{1}{8} \La_i \sxi^* \right) \chi
\]
\be + \left( -i\mu \lbrace \left( \Ga_{1234} + \Ga_{5678} \right),
\La_i\rbrace + \frac{1}{4} \sxi~ \Om_i + \frac{1}{8} \Om_i \sxi \right)
\chi^* = 0,
\label{eqn:diff_chi}
\ee

\noindent where we have used the chirality condition $\Ga_{11}
\vep = \vep$ which implies, for example,
\be
\left( \Ga_{1234} + \Ga_{5678} \right) \Om_i \vep = -i\mu \Ga_+
\Ga_i \vep,
\ee
\noindent and related properties. After some
elementary gamma matrix manipulations, we find that these
constraints become
\be
\left( A_{ij} \Ga_j + \mu^2 \Ga_i - \frac{1}{32} \xi_{jk} \xi_{kl}^*
\Ga_i \Ga_{jl} - \frac{1}{4} \xi_{ij}^* \xi_{jk} \Ga_k \right) \Ga_+
\chi + \frac{i\mu}{8} \Ga_{5678} \Ga_i \sxi ~\Ga_+ \chi^*
= 0,
\ee

\noindent for each $i$, where we have made use of the fact that $\sxi
~\Ga_+ \chi = 0$.  Finally, we substitute explicitly for $\xi_{ij}$,
from the Ansatz (\ref{eqn:ansatz}), giving
\ba
&& \left( A_{ij} \Ga_j + \mu^2 \Ga_i - \frac{1}{8} |\zeta|^2 \Ga_i \left
( i\Ga_{12} + i\Ga_{34} \right) + \frac{1}{2} |\zeta|^2 (\delta_{i1} +
\delta_{i2} )\Ga_i(1 + i\Ga_{12}) \right. \label{eqn:to_solve2}
\nonumber \\
&&\left. + \frac{1}{2} |\zeta|^2 (\delta_{i3} +
 \delta_{i4} ) \Ga_i(1 + i\Ga_{34}) \right) \Ga_+ \chi +
\frac{i}{4}\mu\zeta\, \Ga_{5678} \Ga_i \Ga_{13}(1 + i\Ga_{12})(1 + i\Ga_{34})~\Ga_+ \chi^* = 0.
\label{eqn:to_solve}
\ea

From the above, it is clear that we recover the usual 16 Killing spinors,
annihilated by $\Ga_+$. The question remains which of the cases ($ii$--$iv$)
in (\ref{eqn:dil_soln}) also survive the projections (\ref{eqn:to_solve})
to become supernumerary supersymmetries. Note that cases ($ii$)
and ($iii$) have equal and opposite eigenvalues of the
rotation operators $i\Ga_{12}$ and $i\Ga_{34}$.  The first has
\be
(1+i\Ga_{12}) \chi = 0 = (1-i\Ga_{34}) \chi,
\label{eqn:extra_susy1}
\ee

\noindent whereas the second has
\be
(1-i\Ga_{12}) \chi = 0 = (1+i\Ga_{34}) \chi.
\label{eqn:extra_susy2}
\ee
This leads to a simplification in (\ref{eqn:to_solve}). First,
as these spinors satisfy
\be
\sxi ~\Ga_+ \chi^*=0.
\label{eqn:condition}
\ee

\noindent the last term proportional to $\chi^*$ is eliminated.
Further, the term proportional to $(i\Ga_{12}+i\Ga_{34})$ also vanishes.
Thus (\ref{eqn:to_solve}) reduces to
\be
\left( A_{ij} \Ga_j + \mu^2 \Ga_i + \frac{1}{2} |\zeta|^2
(\delta_{i1} +\delta_{i2} )\Ga_i (1 + i\Ga_{12})
+ \frac{1}{2} |\zeta|^2 (\delta_{i3} +
 \delta_{i4} ) \Ga_i(1 + i\Ga_{34})\right) \Ga_+ \chi = 0.
\ee

\noindent The spinors ($ii$) satisfying (\ref{eqn:extra_susy1}) must
have
\be
A_{ij} = -\mu^2 \delta_{ij} \qquad i,j \ne 1,2, \qquad A_{11} = A_{22}
= -(\mu^2 + |\zeta|^2),
\label{eqn:numcase}
\ee

\noindent whereas in case ($iii$) with (\ref{eqn:extra_susy2}), requires
instead
\be
A_{ij} = -\mu^2 \delta_{ij} \qquad i,j \ne 3,4, \qquad A_{33} = A_{44}
= -(\mu^2 + |\zeta|^2).
\ee

\noindent Thus, although we can have either one or the other, we cannot
have both cases ($ii$) and ($iii$) as supernumerary supersymmetries.
Taking case ($ii$) to be Killing, the metric is thus:
\be
\d s^2 = 2 \d u \d v - \left( \mu^2 \sum_{i=3}^8 x^i x^i + (\mu^2 +
|\zeta|^2) |z^1|^2 \right) \d u^2 + \d s^2(\mathbb{E}^8).
\ee

\noindent Note that this does indeed solve the equation of motion
(\ref{eqn:trace_A}), with $F_5$ and $G_3$ as in (\ref{eqn:ansatz}).

Finally we should consider case ($iv$) in (\ref{eqn:dil_soln}).
Explicitly substituting the corresponding eigenvalues for $i\Ga_{12}$
and $i\Ga_{34}$ into (\ref{eqn:to_solve}) yields
\be
\left( A_{ij} \Ga_j + \mu^2 \Ga_i - \frac{1}{4} |\zeta|^2\Ga_i
\right) \Ga_+ \chi+
i\mu\zeta\, \Ga_{5678} \Ga_i \Ga_{13}~\Ga_+ \chi^* = 0.
= 0.
\label{eqn:newr}
\ee
While this equation has new nontrivial solutions for an appropriate
choice of $A_{ij}$, we have already committed ourselves to that
given in (\ref{eqn:numcase}).
Given this form of the metric, it is straightforward to show that
the eight constraints (\ref{eqn:newr}) cannot be simultaneously satisfied.
For example, consider multiplying the $i=1$ component by $\Ga_{-1}$
\be
- \frac{5}{2} |\zeta|^2 \chi+
2i\mu\zeta\, \Ga_{135678} \chi^*= 0,
\label{eqn:newra}
\ee
using $\Ga_-\chi=0$ for these particular spinors. Similarly
the $i=3$ component yields
\be
- \frac{1}{2} |\zeta|^2 \chi+
2i\mu\zeta\, \Ga_{135678}  \chi^* = 0.
\label{eqn:newrb}
\ee
Hence the difference of these two equations leaves $\chi=0$.
Therefore, all told,
we have a solution which preserves 20 supersymmetries, which includes
the standard 16 supersymmetries of case ($i$) and the 4 supernumerary Killing
spinors of case ($ii$).  All of the 20 supersymmetries which this
specific solution preserves depend on the null coordinate $u$ since,
from (\ref{eqn:diff_chi}) we have
\be
\frac{\d \chi}{\d u} = -\frac{i}{2}\mu \left( \Ga_{1234} + \Ga_{5678}
\right) \chi - \frac{1}{8} \sxi~\chi^*,
\ee

\noindent which one may verify always has a nontrivial solution.

In the following section, we will see that precisely this pp--wave can be
obtained as a Penrose limit of the supergravity solution exhibited by
Pilch and Warner in ref.~\cite{PW1}.


\sect{Taking the Penrose limit at the IR fixed point}
\label{sec:IR}

The solution of five--dimensional $\N=8$ gauged supergravity presented
in ref.~\cite{FGPW} interpolates between two supersymmetric critical
points of the scalar potential.  In the ultra--violet (UV), it gives
the standard maximally supersymmetric AdS$_5$ critical point, and in
the IR it gives the $\N = 2$ AdS$_5$ critical point found in
ref.~\cite{KPW}.  The solution provides a gravity dual of $\N=4$ SYM
theory perturbed by a mass term for a single $\N=1$ chiral
superfield~\cite{KLM,FGPW}, which we take to be $\Phi_3$.  It thus
describes the $\N=1$ RG flow between the $\N=4$ theory and the IR
fixed point, which is a large $N$ limit of the superconformal $\N=1$
theory of Leigh and Strassler~\cite{LS}.  On the gravity side, one has
a solution which preserves 1/8 of the supersymmetries everywhere, this
being enhanced to 1/4 at the IR fixed point.

Since the five--dimensional theory is believed to be a consistent
truncation of type IIB supergravity~\cite{KPW,PW3}, one can lift this
solution directly to ten dimensions.  The resulting $\N=1$
geometry~\cite{PW1,PW2} interpolates between AdS$_5 \times S^5$ in the
UV, and a warped product of another AdS$_5$ with a squashed
five--sphere in the IR.  Before turning to the flow geometry, let us
first restrict ourselves to the IR fixed point solution of
ref.~\cite{PW1}, which is the gravity dual of the $\N=1$
superconformal theory in its own right, {\it i.e.}, it does not
represent the process of deforming away from the UV $\N=4$ fixed
point gauge theory.

For completeness, and to fix our conventions, we present this fixed
point solution in appendix~A.  As discussed therein, we want to work
in coordinates for which the $U(1)_R$ symmetry is simplest so we shift
the $S^3$ Euler angle $\beta \ra \beta + 2\p$, to give a solution with a
global $U(1)_R = U(1)_{\p}$ symmetry.  Performing this coordinate
transformation on the solution (\ref{eqn:app_metric}),
(\ref{eqn:app_5-form}), (\ref{eqn:app_3-form}) and writing the AdS
space in global coordinates gives \ba \d s^2 &=& L^2 \Omega^2 \left(
  -\cosh^2\rho~ \d \t^2 + \d \rho^2 +
  \sinh^2 \rho~ \d \Om^2_3 \right) + \d s^2_5, \label{eqn:PW1} \\
\d s^2_5 &=& \frac{2}{3} L^2 \Om^2 \left[ \d\th^2 + \frac{4 \cos^2
    \th}{(3- \cos 2\th)} (\s_1^2 + \s_2^2) +
  \frac{4 \sin^2 (2\th)}{(3- \cos 2\th)^2} \left( \s_3 + \d \p \right)^2 \right. \nonumber \\
&& \left.  \qquad \qquad \qquad \qquad + \frac{8}{3} \left(
    \frac{2\sin^2 \th - \cos^2 \th}{3-\cos(2\th)} \right)^2 \left(
    \d\p - \frac{2\cos^2 \th}{2\sin^2 \th - \cos^2 \th}
    \s_3 \right)^2 \right], \label{eqn:PW2} \\
F_5 &=& -\frac{2^{5/3}}{3} L^4 \cosh \rho~ \sinh^3 \rho \left( 1 +
  \star \right) \d \t \wedge \d \rho \wedge \ep (S^3), \\
G_3 &=& -i L_0^2 \cos \th \left[ \d \th \wedge \d \p - \frac{8 \cos ^2
    \th}{(3-\cos(2\th))^2} \d \th \wedge \left( \s_3 + \d \p
  \right) \right. \nonumber \\
&& \left. \qquad \qquad \qquad \qquad \qquad \qquad \qquad \qquad -
  \frac{2i \sin (2\th)}{(3-\cos (2\th))} \s_3 \wedge \d \p \right]
\wedge \left( \s_1 + i \s_2 \right). \label{eqn:PW3} \ea

\noindent where
\be
\Om^2 = \frac{2^{1/3}}{\sqrt{3}} \sqrt{3-\cos(2\th)},
\ee

\noindent and the AdS radius, $L$, is given in terms $L_0$, the AdS radius
of the UV spacetime, by
\be
L=\frac{3}{2^{5/3}} L_0.
\ee

\noindent
Note that the three--form field strength $G_3$ could include an
arbitrary constant phase, which we have set to $-1$ here.  The global
isometry group of the metric is $SU(2) \times U(1)_{\beta} \times
U(1)_{\p}$, where $U(1)_{\beta}$ denotes the shift in the Euler angle
$\beta$, rotating $\s_1$ into $\s_2$.  However, from the three--form
one sees that the $U(1)_R$ R--symmetry of the solution as a whole is
$U(1)_R = U(1)_{\p}$, as required.

Since we are working in global coordinates, we can consider the simple
null geodesics for which $\rho=0$.  An examination of the $\th$
geodesic equation shows that one can also consistently set either
$\th=0$ or $\th=\pi/2$.  As discussed in appendix A, the moduli space
of a D3-brane probe in this geometry~\cite{JLP1,JLP2} corresponds to
$\th=0$, whereas the massive direction away from the moduli space
corresponds to $\th=\pi/2$.  We will consider taking the Penrose limit
along a null geodesic in the moduli space, with $\th=0$.  We do also
consider the other class of geodesics, with $\th=\pi/2$, but since we
have not been able to see the relevance of the latter to the gauge
theory, we have consigned the analysis of this case to appendix~B.
Suffice it to say here that we do not understand the significance in
the gauge theory of considering geodesics with angular momentum in the
massive direction, since it is precisely this direction which is to be
``integrated out'' in the ${\cal N}=1$ gauge theory
at the IR fixed point; motion in this direction
(and hence gauge theory operator excitations) should
be effectively frozen in the low energy field theory.

We thus take $\th=0$, in which case the
effective Lagrangian is \be \L = L^2 \Om_0^2 \left[ -\dot{\t}^2 +
  \frac{1}{3} \left( \dot{\alpha}^2 + \sin^2 \alpha \dot{\ga}^2
  \right) + \frac{4}{9} \left( \dot{\p} + \dot{\beta} + \cos \alpha
    \dot{\ga} \right)^2 \right], \ee
\noindent where $\Om_0^2 = 2^{1/3} \sqrt{2/3}$ and a dot denotes
differentiation with respect to the affine parameter.  We can thus also
consider geodesics for which $\alpha=0$, giving\footnote{Although more
general geodesics could be considered here, we suspect that other
geodesics would simply give a Hamiltonian with an alternate linear
combination of the charges $J$ and $J_3$ (to be defined below).}
\be
\L = L^2 \Om_0^2 \left( -\dot{\t}^2 + \frac{4}{9} \dot{\psi}^2
\right),
\ee

\noindent where we have defined
\be
\psi = \p + \beta + \ga,
\label{eqn:psi}
\ee

\noindent to be the direction in which our geodesics
have an angular momentum.  The natural light--cone coordinates are then
\be
u = \frac{1}{2E} \left( \t + \frac{2}{3} \psi \right), \qquad v =
- E L^2 \Om_0^2 \left( \t - \frac{2}{3} \psi \right),
\ee

\noindent where $E$ is the conserved energy associated with the
Killing vector $\del/\del\t$.  If $h$ is the
conserved angular momentum associated with the Killing vector
$\del/\del\psi$, we have $E=(2/3)h$.  We implement the fact that we
are considering $\rho=\th=\alpha=0$ geodesics by taking
\be
\rho = \frac{r}{L}, \qquad \th = \frac{y}{L}, \qquad \alpha = \frac{w}{L},
\ee

\noindent and considering the $L \ra \infty$ limit.

Dropping terms of $\O(1/L^2)$, defining two new angular coordinates as
\be
\hat{\p} = \p - \frac{1}{3} \psi, \qquad \hat{\ga} = \ga - \frac{2}{3}
\psi,
\label{eqn:new_angles}
\ee

\noindent and rescaling $r$, $y$ and $w$, the metric becomes
\be
\d s^2 = 2 \d u \d v - E^2 \left( r^2 + w^2 + 4 y^2 \right) \d u^2 +
\d r^2 + r^2 \d \Om^2_3 + \d y^2 + y^2 \d \hat{\p}^2 + \d w^2 + w^2 \d \hat{\ga}^2.
\label{thewave}
\ee

\noindent With the same definitions of $\hat{\p}$ and $\hat{\ga}$, and
rescalings of coordinates, taking the Penrose limit of the form fields gives
\ba
F_5 &=& - E \left( 1 + \star \right) \d u \wedge  \ep(\mathbb{E}^4), \nonumber \\
G_3 &=& - \sqrt{3} E ~e^{i\beta} ~\d u \wedge \left( \d y - i y^2 \d
\hat{\p} \right) \wedge \left( \d w - i w^2 \d \hat{\ga} \right)
\label{wavefields} \\
&=& \sqrt{3} E ~\d u \wedge \d z^1 \wedge \d z^2 \nonumber,
\ea

\noindent where $\ep(\mathbb{E}^4) = r^3 \d r \wedge \ep(S^3)$ and we have
defined complex coordinates on the remaining $\mathbb{E}^4$.  It
should be clear that the resulting solution is precisely the 5/8
supersymmetric pp--wave considered in the previous section, with
$\mu=-E$ and $\zeta=\sqrt{3}E$.  The Penrose limit we have considered
thus enhances supersymmetry from 1/4 to 5/8.

This is a genuinely new background, and we have a specific
identification of it as dual to a known four--dimensional gauge theory.
Given that this is so, we now turn to an analysis of string theory in
this background.  It should be clear that not all of the worldsheet
scalars will have equal masses, as is the standard
case~\cite{metsaev}: the factor of 4 in the matrix $A_{ij}$ of the
metric (\ref{thewave}) implies that, whereas six of the scalars will
have a mass proportional to $E$, the remaining two will have a mass
proportional to $2E$. This will have non--trivial consequences for the
operator spectrum of the gauge theory as well, as we shall see.


\sect{String propagation}
\label{sec:spectrum}

\subsection{World--sheet analysis: bosonic sector}
\label{sec:bosonic}

The fields which will contribute to our discussion of the world--sheet
bosons are the NS--NS fields, {\it i.e.,} the metric and the
antisymmetric tensor field $B_2$. There are a number of useful
  choices for a gauge within which to work with the $B$--field, with
  (of course) the same resulting physics.  A convenient choice for our
purposes is
\be
B_2 = -\sqrt{3}E ( x^1\d u\wedge \d x^3- x^2\d u\wedge \d x^4).
\ee

\noindent The relevant part of the world--sheet action is:
\begin{eqnarray}
S_B &=& -\frac{1}{4\pi\alpha^\prime}\int \d \sigma \d \tau
\biggl\{\sqrt{-g}g^{\alpha
\beta}(2\partial_{\alpha}U\partial_{\beta}V +
A_{ij}X^iX^j\partial_{\alpha}U\partial_{\beta}U + \partial_{\alpha}X^i\partial_{\beta}X^i)
\nonumber\\
&&\hskip5cm- 2\sqrt{3} E \epsilon^{\alpha
\beta}(X^1\partial_{\alpha}U\partial_{\beta}X^3-X^2\partial_{\alpha}U\partial_{\beta}X^4)
\biggr\},
\end{eqnarray}
where $\epsilon^{01}=1$ and we shall use the familiar world--sheet
gauge choice $g_{\alpha \beta}=\eta_{\alpha \beta}$. We have used
world--sheet coordinates $\sigma^\alpha$, where $\alpha,\beta=0,1$,
and $\sigma^0=\tau, \sigma^1=\sigma$. $A_{ij}$ may be read off from
(\ref{thewave}).

Variation of $V$ gives rise to the equation of motion for
$U$, namely $\nabla^2 U=0$.  So we can work in the standard light--cone gauge with
$U=\al'\pp\tau+{\rm const}$.  In that case,
the worldsheet scalars obey the following equations:
\begin{eqnarray}
&& \nabla^2 X^1-4M^2 X^1+\sqrt{3}M\partial_\sigma X^3 =0,\nonumber\\
&& \nabla^2 X^2-4M^2 X^2-\sqrt{3}M\partial_\sigma X^4 =0,\nonumber\\
&& \nabla^2 X^3-M^2 X^3-\sqrt{3}M\partial_\sigma X^1 =0,\\
&& \nabla^2 X^4-M^2 X^4+\sqrt{3}M\partial_\sigma X^2 =0,\nonumber\\
&& \nabla^2 X^p-M^2 X^p=0,\nonumber
\end{eqnarray}
where $p,q = 5,6,7,8$ will label the directions which are unaffected
by the $B$--field, and where we have set $M = E \al' p^+$.  The
structure of these equations is interesting: they are roughly familiar
from other pp--wave systems (see for
example refs.~\cite{forgacs,RT,metsaevtseytlin}, which follow on from
refs.~\cite{Amati,horror1,horror2,jofre}), but there are crucial
differences brought on by the asymmetry between the 1--3 plane and the
2--4 plane (visible in (\ref{thewave})), which will produce an amusing
mass splitting in the spectrum, as we shall see.  The two independent
components of the standard constraint from world--sheet
reparameterisations, $T_{\alpha\beta}=0$, are \ba
\del_\s V &=& -\frac{1}{\al' p^+} \del_\t X^i \del_\s X^i, \\
\del_\t V &=& -\frac{1}{2 \al' p^+} \left( \del_\t X^i \del_\t X^i +
  \del_\s X^i \del_\s X^i + (\al' p^+)^2 A_{ij} X^i X^j \right), \ea

\noindent which allow for the elimination of $V$ in the usual way.
Integrating the former over $\s$ gives
\be
\int_0^{2\pi} \d \s\,\, \del_\t X^i \del_\s X^i = 0.
\label{eqn:constraint}
\ee

\noindent In the light--cone gauge, the action becomes
\begin{eqnarray}
S_B &=& -\frac{1}{4\pi\alpha^\prime}\int \d \sigma \d \tau
\biggl\{-2\al'\pp\partial_\tau V - (\al'p^+)^2A_{ij}X^iX^j + \eta^{\alpha \beta}\partial_{\alpha}X^i\partial_{\beta}X^i
\nonumber\\
&&\hskip5cm -2\sqrt{3} M \left(X^1\partial_\sigma X^3-X^2\partial_\sigma X^4\right)
\biggr\},
\end{eqnarray}
from which it is easy to derive the Hamiltonian:
\begin{eqnarray}
H_B &=& \frac{1}{4\pi\alpha^\prime}\int_0^{2\pi} \d\sigma
\biggl\{(2\pi\al')^2\Pi^i\Pi^i-(\al'p^+)^2A_{ij}X^iX^j+\partial_\sigma X^i\partial_\sigma X^i
\nonumber\\
&&\hskip5cm -2\sqrt{3} M \left(X^1\partial_\sigma X^3-X^2\partial_\sigma X^4\right)
\biggr\},
\label{eqn:hamiltonian}
\end{eqnarray}
where the conjugate variable to $X^i$ is
\be
\Pi^i=\frac{1}{2\pi\al'}\partial_\tau X^i.
\ee

To solve for the eigenmodes of the system, subject to
the usual periodic boundary conditions $X^i (\t,\s+2\pi) = X^i
(\t,\s)$, we Fourier expand
\be
X^i(\tau,\sigma)=\sum_n C_n^i e^{i\left( \omega_n\t + n
\sigma\right)},
\label{eqn:fourier}
\ee
from which we get the following system of equations:
\begin{eqnarray}
&&  [-\omega^2_n+(n^2+4M^2)]C_n^1-in\sqrt{3}M C_n^3 = 0,\nonumber\\
&&  [-\omega^2_n+(n^2+4M^2)]C_n^2+in\sqrt{3}M C_n^4 = 0,\nonumber\\
&&  [-\omega^2_n+(n^2+M^2)]C_n^3+in\sqrt{3}M C_n^1 = 0,\\
&&  [-\omega^2_n+(n^2+M^2)]C_n^4-in\sqrt{3}M C_n^2 = 0,\nonumber\\
&&  [-\omega^2_n+(n^2+M^2)]C_n^p = 0,\nonumber
\end{eqnarray}
for some unknown coefficients $C_n^i$.  For the $X^p$, the normal
modes are
\begin{equation}
\omega_n^2=n^2+M^2.
\label{basicmodes}
\end{equation}
As is by now well--known, a key feature of this spectrum is that even
the zero modes ($n=0$) have an oscillator frequency $\omega_0=M$
set by the pp--wave background, corresponding to the mass of the
world--sheet bosons associated with those directions.  Beyond that,
there are simply four independent towers of oscillators
(one for each direction) with the mode expansion for these coordinates
being~\cite{metsaevtseytlin}
\be
X^p (\t,\s) = \cos M\t ~x^p_0 + \frac{\al'}{M} \sin M\t ~ p^p_0 + i
\sqrt{\frac{\al'}{2}} \sum_{n \ne 0} \frac{1}{\om_n} \left( \al^p_n
e^{in\s} + \td{\al}^p_n e^{-in\s} \right) e^{-i \om_n\t},
\ee

\noindent where, to ensure reality, we have
\be
\om_n = \sqrt{n^2 + M^2} \qquad (n>0), \qquad \om_n = -\sqrt{n^2 +
M^2} \qquad (n<0),
\ee

\noindent and
\be
\left( \al^p_n \right)^\dagger = \al^p_{-n}, \qquad \left
( \td{\al}^p_n \right)^\dagger = \td{\al}^p_{-n}.
\ee

\noindent The non--vanishing Poisson brackets are easily found to be
\be
[x^p_0,p^q_0]_{PB} = \delta^{pq}, \qquad [\al^p_m,\al^q_n]_{PB} =
[\td{\al}^p_m,\td{\al}^q_n]_{PB} = -i \om_m \delta_{m+n,0} \delta^{pq}.
\label{eqn:poisson1}
\ee

Turning to the directions $i=1,2,3,4$, we have a more complicated
system, since the masses in the 1 and 3, and 2 and 4, directions are
different. We do not expect a simple symmetric result for the coupled
system.  After a bit of algebra, we obtain the intriguing formula:
\begin{equation}
\omega_n^2=\frac{1}{2}\left(2n^2+5M^2\pm\sqrt{12n^2M^2+9M^4}\right)
\equiv \left(\om_n^{\pm}\right)^2,
\label{moremodes}
\end{equation}
for these directions. This will give {\it
  two} distinct pairs of frequencies, giving again four independent
families of oscillators. We observe that the natural
frequencies of the zero modes are $\omega_0^-=\omega_0=M$ and $\omega_0^+=2M$, as
  expected.  Explicitly, the mode expansions for these coordinates are thus
\ba
X^1 (\t,\s) &=& \cos 2M\t ~x^1_0 + \frac{\al'}{2M} \sin 2M\t ~p^1_0 + i
\sqrt{\frac{\al'}{2}} \sum_{n \ne 0}  \left[ \frac{1}{\om_n^+} \left( \beta^1_n
e^{in\s} + \td{\beta}^1_n e^{-in\s} \right) e^{-i \om_n^+ \t}
\right. \nonumber \\
&& \hskip6.5cm \left. + \frac{1}{\om_n^-} \left( \ga^1_n
e^{in\s} + \td{\ga}^1_n e^{-in\s} \right) e^{-i \om_n^- \t} \right],
\\
X^3 (\t,\s) &=& \cos M\t ~x^3_0 + \frac{\al'}{M} \sin M\t ~p^3_0 + i
\sqrt{\frac{\al'}{2}} \sum_{n \ne 0}  \left[ \frac{c_n^+}{\om_n^+} \left( \beta^1_n
e^{in\s} - \td{\beta}^1_n e^{-in\s} \right) e^{-i \om_n^+ \t}
\right. \nonumber \\
&& \hskip6.5cm \left. + \frac{c_n^-}{\om_n^-} \left( \ga^1_n
e^{in\s} - \td{\ga}^1_n e^{-in\s} \right) e^{-i \om_n^- \t} \right],
\ea

\noindent where $\om_n^{\pm}$ is given by the positive (negative) root
of (\ref{moremodes}) for positive (negative) $n$ and
\be
c_n^{\pm} = \frac{i}{2\sqrt{3}nM} \left( -3M^2 \pm \sqrt{12n^2M^2+9M^4} \right),
\ee

\noindent which obeys $c_n^+ c_n^- =1$.  Similar expressions hold for $X^2$ and $X^4$, with
$\{\beta^1,\ga^1,\td{\beta}^1,\td{\ga}^1\}$ replaced by
$\{\beta^2,\ga^2,\td{\beta}^2,\td{\ga}^2\}$ and $c_n^{\pm}$
replaced by $-c_n^{\pm}$.  With $A=1,2$, reality of the coordinates implies
\be
\left( \beta^A_n \right)^\dagger = \beta^A_{-n}, \qquad \left
( \ga^A_n \right)^\dagger = \ga^A_{-n}, \qquad \left
( \td{\beta}^A_n \right)^\dagger = \td{\beta}^A_{-n}, \qquad \left
( \td{\ga}^A_n \right)^\dagger = \td{\ga}^A_{-n}.
\ee

\noindent In addition to (\ref{eqn:poisson1}), the remaining
non--vanishing Poisson brackets are then
\ba
&& [x^A_0,p^B_0]_{PB} = [x^{A+2}_0,p^{B+2}_0]_{PB} = \delta^{AB},
\nonumber \\
&& [\beta^A_m, \beta^B_n]_{PB} = [\td{\beta}^A_m, \td{\beta}^B_n]_{PB} = -i \om_m^+
\frac{c_m^-}{c_m^- - c_m^+} \delta_{m+n,0} \delta^{AB}, \\
&& [\ga^A_m, \ga^B_n]_{PB} =
[\td{\ga}^A_m, \td{\ga}^B_n]_{PB} = -i \om_m^- \frac{c_m^+}{c_m^+ -
c_m^-} \delta_{m+n,0} \delta^{AB}, \nonumber
\ea

The constraint (\ref{eqn:constraint}) becomes $N=\td{N}$, where
\ba
N &=& \sum_{n\ne 0} n \left[ \frac{1}{\om_n} \al^p_{-n} \al^p_n +
\frac{1}{\om_n^+} \left( 1 - c_n^{+2} \right) \beta^A_{-n} \beta^A_n  +
\frac{1}{\om_n^-} \left( 1 - c_n^{-2} \right) \ga^A_{-n} \ga^A_n
\right], \nonumber \\
\td{N} &=& \sum_{n\ne 0} n \left[ \frac{1}{\om_n} \td{\al}^p_{-n} \td{\al}^p_n +
\frac{1}{\om_n^+} \left( 1 - c_n^{+2} \right) \td{\beta}^A_{-n} \td{\beta}^A_n  +
\frac{1}{\om_n^-} \left( 1 - c_n^{-2} \right) \td{\ga}^A_{-n} \td{\ga}^A_n
\right],
\label{eqn:number_ops}
\ea

\noindent and the Hamiltonian (\ref{eqn:hamiltonian}) is
\[\hskip-7cm
H_B = \frac{1}{2\al'} \left( \al'^2 p^i_0p^i_0 + 4M^2 \sum_{i=1,2} x^i_0 x^i_0 + M^2 \sum_{i=3}^8 x^i_0
x^i_0  \right) \nonumber
\]
\be
+ \frac{1}{2} \sum_{n \ne 0} \left( \al^p_{-n} \al^p_n
+ \td{\al}^p_{-n} \td{\al}^p_n + \left( 1 - c_n^{+2} \right) \left( \beta^A_{-n}
\beta^A_n + \td{\beta}^A_{-n} \td{\beta}^A_n \right) + \left( 1 - c_n^{-2} \right) \left( \ga^A_{-n}
\ga^A_n + \td{\ga}^A_{-n} \td{\ga}^A_n \right) \right).
\label{eqn:hamiltonian2}
\ee

To quantise the system, we replace the Poisson brackets with
commutators in the usual way.  We further take, for $n>0$,
\[
\al^p_n = \sqrt{\om_n} ~a^p_n, \qquad \al^p_{-n} = \sqrt{\om_n}
~\bar{a}^p_n,
\]
\ba
&& \beta^A_n = \sqrt{\om_n^+
\frac{c_n^-}{c_n^- - c_n^+}}
~b^A_n, \qquad \beta^A_{-n} = \sqrt{\om_n^+
\frac{c_n^-}{c_n^- - c_n^+}} ~\bar{b}^A_n, \\
&& \ga^A_n = \sqrt{\om_n^- \frac{c_n^+}{c_n^+ -
c_n^-}}
~c^A_n, \qquad \ga^A_{-n} = \sqrt{\om_n^- \frac{c_n^+}{c_n^+ -
c_n^-}} ~\bar{c}^A_n, \nonumber
\ea

\noindent and similarly for the the independent set of operators with a
tilde, and combine the zero modes as
\ba
&& a_0^i = \frac{1}{\sqrt{4M \al'}} \left( \al' p^i_0 - 2iM x^i_0 \right), \qquad
\bar{a}_0^i = \frac{1}{\sqrt{4M \al'}} \left( \al'p^i_0 + 2iM x^i_0 \right),
\qquad (i = 1,2), \\
&& a_0^i = \frac{1}{\sqrt{2M \al'}} \left( \al' p^i_0 - iM x^i_0 \right), \qquad
\bar{a}_0^i = \frac{1}{\sqrt{2M \al'}} \left( \al' p^i_0 + iM x^i_0 \right),
\qquad (i = 3, \ldots, 8).
\ea

\noindent The new creation and annihilation operators obey the
standard harmonic oscillator commutation relations
\be
[a_0^i,\bar{a}_0^j]=\delta^{ij}, \qquad [a_m^p,\bar{a}_n^q] =
\delta_{mn}\delta^{pq}, \qquad [b_n^A,\bar{b}_m^B] =
[c_n^A,\bar{c}_m^B] = \delta_{nm} \delta^{AB},
\ee

\noindent and similarly for the tilded set of operators.  In this basis, the Hamiltonian
(\ref{eqn:hamiltonian2}) becomes
\be
H = \Delta E + 2M \sum_{i=1,2} N_0^{(i)} + M \sum_{i=3}^8 N_0^{(i)} +
\sum_{n > 0} \left( \omega_n N_n^{(a)} + \om_n^+
N_n^{(b)} + \om_n^- N_n^{(c)} \right),
\label{eqn:hamiltonian3}
\ee

\noindent where $\omega_n$ and $\om_n^{\pm}$ are given by
(\ref{basicmodes}) and (\ref{moremodes}) respectively and $\Delta E$ is
the zero point energy.  The occupation numbers are given by
\be
N_n^{(a)} = \bar{a}^p_{n} a^p_n + \td{\bar{a}}^p_{n} \td{a}^p_n,
\ee

\noindent and similarly for $N_n^{(b)}$ and $N_n^{(c)}$, and we have defined
\be
N_0^{(i)} = \bar{a}_0^i a_0^i.
\ee

\noindent The spectrum
of the bosonic string is thus of the same form as in the maximally
supersymmetric case~\cite{metsaev,BMN,metsaevtseytlin}, the only
difference being the more complicated frequencies $\om_n^{\pm}$.  We
will see shortly that precisely the same frequencies appear in the
normal modes of some of the fermions.

At this point, however, the interesting structure of our equations
leads us to consider a slightly more general class of scenarios in
appendix~C. There, we comment briefly on an interesting property of
the equations of motion for the directions affected by the $B$--field,
pointing to the possibility of a new type of stringy instability.

\subsection{World--sheet analysis: fermionic sector}

The problem as to how to include R--R fields in the worldsheet
analysis of the superstring is a difficult one.  Techniques utilising
coset superspaces have been used in an attempt to construct actions
for superstrings in AdS backgrounds (\emph{e.g.} ref.~\cite{MT}),
although the resulting action is difficult to quantise explicitly.
More recently, such techniques have been applied~\cite{metsaev,BMN} to
the maximally supersymmetric pp--wave background of ref.~\cite{BFHP1}.  In
the light--cone gauge, the superstring action in this background
simplifies considerably and can, in fact, be quantised: it turns out
the five--form field strength only gives rise to mass terms for the
fermions.\footnote{Other techniques can be used to derive the relevant
  action~\cite{CLP2}: since the eleven--dimensional supermembrane
  action is known to $\O(\th^2)$~\cite{WPP}, dimensional reduction
  will give rise~\cite{CLPS} to the superstring action to the same
  order in the fermions; and this is all that is required in the case
  at hand.}  More heuristically, since this background admits a null
Killing vector, it can be argued~\cite{metsaevtseytlin} that the
fermionic action is a direct covariantisation of the flat action, at
least in the standard light--cone gauge.

In our conventions, the light--cone gauge is implemented \emph{via}
\be
\Ga^- \theta = \Ga_+ \theta = 0,
\ee

\noindent in which case the fermionic action is
simply~\cite{metsaevtseytlin}
\be
S_F = \frac{i}{\pi} \int \d \s \d \t \left( \eta^{\alpha \beta} \delta_{IJ}
- \ep^{\alpha \beta} \rho_{IJ} \right) \del_{\alpha} X^a \del_{\beta} X^b ~\bar{\th}^I ~\Ga_a
\D_b \th^J,
\ee

\noindent where $I,J = 1,2$ denote the two 16--component Majorana--Weyl
spinors.  In terms of the Pauli matrices, $\t_i$, the two--dimensional
gamma matrices are $\rho^0 = i\t_2$ and $\rho^1 = \t_1$, so
that $\rho = \rho^0 \rho^1 = \t_3$.  With $G_3 = H_3 + i F_3$, and
viewed as acting on a column matrix, the supercovariant derivative
then takes the form
\be
\D_a = D_a - \frac{1}{96} \not{\!\!H}_a \rho - \frac{1}{96}\not{\!F}_a \rho_1 +
\frac{1}{480} F_{b_1 \ldots b_5} \Ga^{b_1 \ldots b_5} \Ga_a \rho_0,
\ee

\noindent where
\be
\not{\!\!H}_a = H_{bcd} \Ga_a^{~bcd} - 9 H_{abc} \Ga^{bc},
\ee

\noindent and likewise for $\not{\!F}_a$.  In the light--cone gauge,
the action simplifies considerably, and we have~\cite{metsaevtseytlin,CLP2}
\ba
&& S_F = - \frac{i}{\pi} \al' p^+ \int \d \s \d \t \left\{
\bar{\th} ~ \Ga_- \left( \del_\t \th + \rho ~\del_\s \th \right) +
\frac{1}{8} \al' p^+ \bar{\th} ~ \Ga_- \not{\!\!H} \rho
~\th + \frac{1}{8} \al' p^+ \bar{\th} ~\Ga_- \not{\!F}_3 \rho_1 ~\th
\right. \nonumber \\
&& \hskip8cm \left. + \frac{1}{240} \al'
p^+ \bar{\th} ~ \Ga_- \not{\!F}_5 \rho_0 ~\th \right\},
\ea

\noindent where now
\be
\not{\!\!H} = H_{uij} \Ga_{ij}, \qquad \not{\!F}_3 = F_{uij} \Ga_{ij},
\qquad \not{\!F}_5 = F_{uijkl} \Ga_{ijkl}.
\ee

\noindent We should note that the NS--NS three-form gives rise to a
chiral interaction, whereas the R--R three--form field strength gives further mass
terms~\cite{metsaevtseytlin}.

Finally, then, we substitute for $H_{u13} = - H_{u24} = F_{u14} =
F_{u23} = \sqrt{3} E$ and $F_{u1234}= F_{u5678} = -E$, and rewrite in
terms of $\th^1$ and $\th^2$, giving
\ba
&& S_F = -\frac{i}{\pi} \al' p^+ \int \d \s \d \t \left\{
\th^1 \Ga_- \del_+ \th^1 + \th^2 \Ga_- \del_- \th^2 + \frac{\sqrt{3}}{2} M
\th^1 \Ga_- (\Ga_{14}+\Ga_{23}) \th^2  \right. \nonumber \\
&& \hskip1cm
\left. + \frac{\sqrt{3}}{4} M \th^1 \Ga_- (\Ga_{13}-\Ga_{24}) \th^1 -
  \frac{\sqrt{3}}{4} M \th^2 \Ga_- (\Ga_{13}-\Ga_{24}) \th^2 - 2 M
  \th^1 \Ga_- \Ga_{1234} \th^2
\right\},
\ea

\noindent where
\be
\del_{\pm} = \del_{\t} \pm \del_{\s}.
\ee

\noindent The equations of motion for $\th^1$ and $\th^2$ are then
\ba
&& \del_+ \th^1 - M \Ga_{1234} \th^2 + \frac{\sqrt{3}}{4} M (\Ga_{14}
+ \Ga_{23}) \th^2 + \frac{\sqrt{3}}{4} M (\Ga_{13} - \Ga_{24}) \th^1
=  0, \nonumber \\
&& \del_- \th^2 + M \Ga_{1234} \th^1 + \frac{\sqrt{3}}{4} M (\Ga_{14}
+ \Ga_{23}) \th^1 - \frac{\sqrt{3}}{4} M (\Ga_{13} - \Ga_{24}) \th^2
=  0. \label{th-eom}
\ea

\noindent The next step is again to Fourier expand
\[
\th^I(\tau,\sigma)=\sum_n \th^I_n(\tau) e^{in\sigma},
\]

\noindent giving
\ba
&& \dot{\th}^1_n + M \left(\frac{\sqrt{3}}{4} (\Ga_{14} + \Ga_{23}) -
  \Ga_{1234} \right) \th^2_n + \left(\frac{\sqrt{3}}{4} M (\Ga_{13} -
  \Ga_{24}) + in \right) \th^1_n = 0, \nonumber  \\
&& \dot{\th}^2_n + M \left(\frac{\sqrt{3}}{4} (\Ga_{14} + \Ga_{23}) +
  \Ga_{1234} \right) \th^1_n - \left(\frac{\sqrt{3}}{4} M (\Ga_{13} -
  \Ga_{24}) + in \right) \th^2_n = 0. \label{thn-eom}
\ea

Differentiating with respect to $\t$ and using (\ref{thn-eom}) again
to eliminate the first derivatives, results in
\be
\ddot{\vep}_n + A_n \vep_n = 0,
\label{ka-eom}
\ee

\noindent where
\be
A_n = \left(n^2 + \frac{7 M^2}{4}\right) I - \frac{3M^2}{4} \Ga_{1234} -
  \frac{i\sqrt{3} Mn}{2} (\Ga_{13} - \Ga_{24}) - \frac{i3 M^2}{4}
  (\Ga_{12} + \Ga_{34}),
\ee

\noindent and we have re--combined $\th^1$ and $\th^2$ into a single complex spinor
$\vep = \th^1 + i\th^2$.  In order to solve (\ref{ka-eom}) we
need to find the eigenspinors of the matrix $A_n$.  To do this, we consider the constant spinors in the Chevalier
basis as in section \ref{sec:adding_susy}.  In particular, we focus
on the eigenvalues of $i\Gamma_{12}$ and $i\Gamma_{34}$, and denote
the spinors as $\vep^{\pm \pm}$ where
\ba
i \Ga_{12}~ \vep^{\pm (\cdot)} & = & \pm \vep^{\pm (\cdot)},\nonumber \\
i \Ga_{34}~ \vep^{(\cdot) \pm} & = & \pm \vep^{(\cdot) \pm}.
\ea

\noindent so that
\ba
A_n \vep^{++} & = & (M^2 + n^2) \vep^{++} - \sqrt{3}Mn \vep^{--},
\nonumber \\
A_n \vep^{+-} & = & (M^2 + n^2) \vep^{+-}, \nonumber \\
A_n \vep^{-+} & = & (M^2 + n^2) \vep^{-+}, \\
A_n \vep^{--} & = & (4 M^2 + n^2) \vep^{--} - \sqrt{3}Mn \vep^{++}.\nonumber
\ea

\noindent
Therefore, at each level $n$, there are four fermionic oscillators
with frequency given by \be \omega^2_n = n^2 + M^2, \ee

\noindent and there are four fermionic oscillators with frequencies
\be
\left(\om_n^{\pm}\right)^2 = \frac{1}{2}\left(2n^2+5M^2\pm\sqrt{12n^2M^2+9M^4}\right).
\ee

\noindent These exactly match the frequencies (\ref{basicmodes}) and
(\ref{moremodes}) found for the bosonic oscillators above, as
presumably required by supersymmetry.

The mode expansions of the fermions are not particularly enlightening,
and we will not exhibit them here.  Suffice it to say that, given the
above results concerning the frequencies of the fermion modes, we fully
expect the total Hamiltonian to be of the same form as the purely
bosonic result (\ref{eqn:hamiltonian3}).  That is,
\be
H = \Delta E + 2M \sum_{i=1,2} N_0^{(i)} + M \sum_{i=3}^8 N_0^{(i)} +
\sum_{n > 0} \left( \omega_n N_n^1 + \om_n^+
N_n^2 + \om_n^- N_n^3 \right),
\label{eqn:total_H}
\ee

\noindent where, in analogy with ref.~\cite{metsaevtseytlin}, the zero point energy $\Delta E = ( 6 \times 1/2 + 2
\times 2 \times 1/2 ) M = 5 M$, the fermion zero modes appear in
$N_0^{(i)}$ and the level operators $N_n^{1,2,3}$ now also include the
relevant contributions from the fermions.


\sect{Gauge theory}
\label{sec:ft}

Let us first consider the light--cone Hamiltonian
\be
H = i\del_u = \frac{\del \t}{\del u} i\del_{\t} + \frac{\del \p}{\del u}
i\del_{\p} + \frac{\del \beta}{\del u} i\del_{\beta} + \frac{\del \ga}{\del u}
i\del_{\ga},
\ee

\noindent where
\be
\t = Eu-\frac{1}{2 \Om_0^2} \frac{v}{EL^2}, \qquad \psi = \frac{3}{2}
\left( Eu + \frac{1}{2 \Om_0^2} \frac{v}{EL^2} \right), \qquad \p = \hat{\p} +
\frac{1}{3} \psi, \qquad \ga = \hat{\ga} + \frac{2}{3} \psi.
\ee

\noindent Since $U(1)_{\beta}$ is not a symmetry of the gauge theory
superpotential, there is no conserved charge associated with the operator
$i\del_{\beta}$, and so it would not make sense to have this term
present in the Hamiltonian.  Happily, however, we have
\be
\beta = \psi - (\p + \ga) = - (\hat{\p} + \hat{\ga}),
\ee

\noindent so that $\del \beta/\del u = 0$ as required.  The scaling
dimension, $\Delta$, the R--charge, $J$, and the ``flavour'' charge,
$J_3$, associated with the $U(1)_{\ga}$ diagonal subgroup of the
global ``flavour'' $SU(2)$, are given by
\be \Delta = i \del_{\t},
\qquad J = -i\del_{\p}, \qquad J_3 = i\del_{\ga}, \ee

\noindent so that, setting $E=1$ for convenience,
\be
H = \Delta - \frac{1}{2} J + J_3.
\ee

\noindent Likewise, the light--cone momentum is given by
\be
P = i\del_v = -\frac{1}{2 \Om_0^2} \frac{1}{L^2} \left( \Delta +
\frac{J}{2} - J_3 \right).
\ee

\noindent Since both of these quantities should remain fixed after
taking the Penrose limit, in analogy with ref.~\cite{BMN}, we are
interested in operators with large R-- and flavour--charges:
\be
J, J_3 \sim L^2 \sim N^{1/2},
\ee

\noindent as we take the $N \ra \infty$ limit, keeping $g_{YM}^2$
fixed and small.  In this limit of {\it infinite} 't Hooft coupling,
we must further demand that $\Delta - (J/2) + J_3$ is kept fixed, so
that the light--cone Hamiltonian remains finite.

The values of $\Delta, J, J_3$ and $H$ for the complex scalar fields
appearing as the lowest--order components in the expansion of the
three chiral and three anti--chiral
superfields are listed in table \ref{table:bosons}.
Remembering that it is $\Phi_3$ which is massive, and can be
integrated out as $\Phi_3 \sim [\Phi_1, \Phi_2]$, the values of $H$
which we find make sense: the energy of $\p_3$ is equal to the sum
of the energies of $\p_1$ and $\p_2$.

\begin{table}
\begin{center}
\renewcommand{\arraystretch}{1.5}
\begin{tabular}{||l|l|l|l|l||} \hline
               & $\Delta$ & $J$   & $J_3$  & $H$   \\ \hline
$\p_1$       & 3/4      & \phantom{$-$}1/2   & \phantom{$-$}1/2    & 1     \\ \hline
$\p_2$       & 3/4      & \phantom{$-$}1/2   & $-$1/2   & 0     \\ \hline
$\p_3$       & 3/2      & \phantom{$-$}1     & \phantom{$-$}0      & 1     \\ \hline
$\bar{\p}_1$ & 3/4      & $-$1/2  & $-$1/2   & 1/2   \\ \hline
$\bar{\p}_2$ & 3/4      & $-$1/2  & \phantom{$-$}1/2    & 3/2   \\ \hline
$\bar{\p}_3$ & 3/2      & $-$1    & \phantom{$-$}0      &2     \\ \hline
\end{tabular}
\caption{The conformal dimensions, charges and light--cone energies of
the complex scalar fields appearing as the lowest--order components in
the expansions of the three chiral and three anti--chiral superfields.}
\label{table:bosons}
\end{center}
\end{table}

The first prediction from the spectrum found in section
\ref{sec:spectrum} is that there should be a \emph{unique} light--cone
ground state with large $\Delta, J$ and $J_3$.  It is simply that
state for which all the occupation numbers in (\ref{eqn:total_H})
vanish.  This corresponds in the gauge theory to the operator $\Tr
(\p_2^{2J})$.  It has $H=0$ since it is chiral --- its conformal
dimension is simply the na\"{\i}ve value $\Delta = 3J/2$.  The second
prediction (and this is where we depart from the previous results
concerning both the $\N = 4$~\cite{BMN} and $\N = 1$~\cite{IKM,GO,PS}
theories) is that there should be precisely {\it six} bosonic
operators with $H=1$ and {\it two} bosonic operators with $H=2$,
corresponding to the zero modes of the worldsheet scalars.  Four are
straightforward to write down; they are simply derivatives of the
ground state operator \be \Tr (D_k \p_2^{2J}), \qquad (k = 1,2,3,4)\ ,
\ee coming from inserting covariant derivatives along any of the four
spacetime directions (in an Euclidean discussion). This follows
straightforwardly from the descendant of the action of the conformal
group.

Another operator with $H=1$ is $\Tr(\p_1 \p_2^{2J})$.  This
is again chiral and so its conformal dimension is the sum of those of
its constituents.  This leaves a sixth bosonic operator to be found,
which we can look for in analogy with the analysis of Itzhaki \emph{et
al.}~\cite{IKM} in the $T^{1,1}$ case.  We propose
\be
\Tr(\bar{\p}_1 \p_2^{2J}),
\ee

\noindent as our sixth operator with $H=1$.  Since it is not chiral, its conformal dimension is not
necessarily the na\"{\i}ve one found using the values in table
\ref{table:bosons}.

To understand why this operator should have $\Delta = (6J+5)/4$, consider
\be
\Tr(\bar{\p}_1 \p_2).
\ee

\noindent Perusal of the tables in ref.~\cite{FGPW}, shows that this  operator is in
the same $\N=2$ supermultiplet as the conserved $SU(2)$ current and
therefore its dimension is the same as its free--field value,
\emph{i.e.}  $\Delta = 2$, which gives $H=1$.  Unfortunately there
does not seem to be a field theory method to derive the conformal
dimension of $\Tr(\bar{\p}_1 \p_2^{2J})$.  For the equivalent operator
in the $T^{1,1}$ case, Itzhaki \emph{et al.}~\cite{IKM} were able to
find the relevant conformal dimension using a standard AdS/CFT formula
relating the conformal dimension to the Laplacian on $T^{1,1}$.  In
the case of the Pilch--Warner geometry considered herein, we do not
have such a formula and so the proposed conformal dimension of
$\Tr(\bar{\p}_1 \p_2^{2J})$ is somewhat more conjectural than one
might have liked.

We should also consider the two operators with $H=2$, which contribute
at the same ``level'' as the six operators with $H=1$.  However, there
are a host of candidates, and none of them seem to have protected
conformal dimensions.  Indeed, partly for this reason, we have not
been able to satisfactorily identify these operators, and so leave it
as a conjecture that there are precisely two such operators with $H=2$
and large $\Delta, J$ and $J_3$ dual to those seen in the string
theory spectrum.

Given this limited success with the bosonic operators at the lowest
lying levels, let us now turn our attention to the fermionic ones.
{}From table \ref{table:fermions} one can immediately see that the
following operators have $H=1$: \be \Tr(\chi_2 \p_2^{2J}), \qquad
\Tr(\psi \p_2^{2J}).  \ee

\noindent These give four fermionic operators since both $\chi$ and $\psi$ are
two--component Weyl fermions.  The first two are the supersymmetry
variation of the ground state operator.  The second two involve the
gaugino, $\psi$.  The remaining two operators to be found are the
fermionic counterparts of $\Tr(\bar{\p}_1 \p_2)$:
\be
\Tr(\bar{\chi}_1 \p_2^{2J}).
\ee

\noindent Again, we do not consider fermionic operators with $H=2$,
but expect that there are precisely two of them as above.

\begin{table}
\begin{center}
\renewcommand{\arraystretch}{1.5}
\begin{tabular}{||l|l|l|l|l||} \hline
               & $\Delta$ & $J$   & $J_3$  & $H$   \\ \hline
$\chi_1$       & 5/4      & $-$1/2  & \phantom{$-$}1/2    & 2     \\ \hline
$\chi_2$       & 5/4      & $-$1/2  & $-$1/2   & 1     \\ \hline
$\psi$         & 3/2      &  \phantom{$-$}1    & \phantom{$-$}0      & 1     \\ \hline
$\bar{\chi}_1$ & 5/4      &  \phantom{$-$}1/2  & $-$1/2   & 1/2   \\ \hline
$\bar{\chi}_2$ & 5/4      &  \phantom{$-$}1/2  & \phantom{$-$}1/2    & 3/2   \\ \hline
$\bar{\psi}$   & 3/2      & $-$1    & \phantom{$-$}0      & 2     \\ \hline
\end{tabular}
\caption{The conformal dimensions, charges and light--cone energies of
  the gauginos, $\psi$, and the fermionic components, $\chi_1$ and
$\chi_2$, of the chiral superfields $\Phi_1$ and $\Phi_2$, and their
anti--chiral counterparts.  We do
not consider the components of $\Phi_3$ here,
since it should not enter our discussion at all.}
\label{table:fermions}
\end{center}
\end{table}

Of course, our ultimate aim should be to reproduce the form of the
string spectrum (\ref{eqn:total_H}) from the gauge theory, along the
lines of Berenstein, Maldacena and Nastase~\cite{BMN}.  Rewriting this
in gauge theory variables, we would want to derive that, for the
$5,6,7,8$ directions, \be ( \Delta - \hat{J} )_n = \sqrt{1 + \frac{n^2
    g_{YM}^2 N}{\hat{J}^2}}, \ee

\noindent where $\hat{J} = J/2 - J_3$, and we have used the fact that $L^4 = g_{YM}^2 N
\al'^2$ and $P \sim \hat{J}/L^2$.  For the $1,2,3,4$ directions, we
should have the very interesting result
 \be ( \Delta - \hat{J} )_n =
\sqrt{ \frac{5}{2} + \frac{n^2 g_{YM}^2 N}{\hat{J}^2} \pm
  \frac{1}{2} \sqrt{9 + 12 \frac{n^2 g_{YM}^2 N}{\hat{J}^2} }}.
\ee

\noindent We will leave the  task of verifying this
prediction directly from gauge theory for future work. See section~1
for more discussion of the gauge theory implications.

Having considered the IR fixed point geometry and its Penrose limits
in some detail, we now move away from the fixed point, turning to the
flow geometry of ref.~\cite{PW2}.


\sect{Taking the Penrose limit along the flow}
\label{sec:flow}

If we are to take the Penrose limit of the flow geometry away from the
fixed points, we are forced to use ``Poincare'' coordinates on the
``AdS'' space.  Using the the same coordinates on the squashed
five--sphere as in section \ref{sec:IR} above, the flow geometry is described by the metric~\cite{PW2}
\ba
\d s^2 &=& \frac{X^{1/2} \cosh \chi}{\rho} \left( e^{2A} \d s^2(\mathbb{M}^4)
+ \d r^2 \right) + \d s^2_5, \label{eqn:metric1} \nonumber \\
\d s^2_5 &=& L_0^2 \frac{X^{1/2}\mbox{sech} \chi}{\rho^3} \left[ \d\th^2 +
\frac{\rho^6 \cos^2 \th}{X} \left( \s_1^2 + \s_2^2 \right) +
\frac{\rho^{12} \sin^2 (2\th)}{4 X^2} \left( \s_3 +
\frac{(2+\rho^6)}{2\rho^6} \d \p \right)^2 \right. \nonumber \\
&& \qquad \qquad \left. + \frac{\rho^6 \cosh^2
\chi}{4 X^2} ( 2 \sin^2 \th - \cos^2 \th )^2 \left( \d \p -
\frac{2 \cos^2 \th}{(2\sin^2 \th - \cos^2 \th)} ~\s_3
\right)^2 \right],
\label{eqn:metric2}
\ea

\noindent where
\be
X = \cos^2 \th + \rho^6 \sin^2 \th.
\ee

\noindent We are still free to choose geodesics for which $\th = 0$
and $\alpha=0$.  Considering also a constant point on $\mathbb{E}^3$, the
effective Lagrangian is
\be
\L = \frac{\cosh \chi}{\rho} \left( -e^{2A} \dot{t}^2 + \dot{r}^2 +
\frac{L_0^2}{4} \rho^4 \dot{\psi}^2 \right),
\ee

\noindent where $\psi$ is defined in (\ref{eqn:psi}).  As above, there
are two conserved quantities, $E$ and $h$, associated with the Killing
vectors $\del/\del t$ and $\del/\del \psi$ respectively.  The $t$ and
$\psi$ equations then give
\be
\dot{t} = E L_0 \frac{\rho}{\cosh \chi} e^{-2A}, \qquad \dot{\psi} =
\frac{h}{\rho^3 \cosh \chi},
\ee

\noindent and the null condition is
\be
\dot{r} = E L_0 \frac{\rho}{\cosh \chi} \sqrt{ e^{-2A} -
\frac{h^2}{E^2} \frac{1}{4\rho^4} },
\ee

\noindent where we have chosen the arbitrary sign in the above to be
positive.  Of course, we cannot integrate to find $r(\la)$, but
we do not need to.

Following refs.~\cite{BFHP2,BFP}, we introduce coordinates
$\{u,v,x\}$ such that $g_{uu} = 0 = g_{ux}$ and $g_{uv}=1$.  In other
words, just as in (\ref{eqn:coords1}), we have
\ba
\del_u &=& \dot{r} \del_r + \dot{t} \del_t + \dot{\psi} \del_{\psi},
\nonumber \\
\del_v &=& -\frac{1}{EL_0} \del_t, \\
\del_x &=& \frac{1}{L_0} \del_{\psi} + \frac{1}{4} \frac{h}{E} \del_t, \nonumber
\ea

\noindent which gives
\ba
\d r &=& EL_0 \frac{\rho}{\cosh \chi} \left(e^{-2A} -
\frac{h^2}{E^2} \frac{1}{4\rho^4} \right)^{1/2} ~\d u, \nonumber \\
\d t &=& EL_0 \frac{\rho}{\cosh \chi} e^{-2A} \d u -
\frac{\d v}{EL_0} + \frac{1}{4} \frac{h}{E} \d x, \\
\d \psi &=& \frac{h}{\rho^3 \cosh \chi} \d u  + \frac{\d x}{L_0}. \nonumber
\ea

\noindent Substituting for these in the metric (\ref{eqn:metric2}), taking
\be
\th = \frac{y}{L_0}, \qquad \alpha = \frac{w}{L_0},
\ee

\noindent and dropping all terms of $\O (1/L_0)$, we find
\[
\d s^2 = 2\d u\d v + \frac{1}{4} \rho^3 \cosh \chi e^{2A} \left( e^{-2A} -
\frac{h^2}{E^2} \frac{1}{4\rho^4} \right) \d x^2 + \frac{\cosh \chi}{\rho} ~e^{2A} \d s^2
(\mathbb{E}^3) + \frac{{\rm sech} \chi}{\rho^3} \left( \d y^2 + y^2 \d
\hat{\p}^2 \right)
\]
\be
+ \frac{1}{4} {\rm sech} \chi \rho^3 \left( \d w^2 + w^2 \d
\hat{\ga}^2 \right) - \frac{h^2}{4} \cosh \chi ~\left( \rho^3 y^2 +
\frac{1}{4} \frac{w^2}{\rho^3} \right) \d u^2,
\ee

\noindent where
\be
\d \hat{\p} = \d \p - \rho^6 \sinh^2 \chi \frac{\d \psi}{2}, \qquad
\d \hat{\ga} = \d \ga - \cosh^2 \chi \frac{\d \psi}{2}.
\ee

\noindent Note that in the IR, these reduce to the angular variables
in (\ref{eqn:new_angles}) as required.

To write this in terms of Brinkman coordinates, define
\[
E(u) = \frac{1}{2} \rho^{3/2} \cosh^{1/2} \chi e^A \sqrt{e^{-2A} -
\frac{h^2}{E^2} \frac{1}{4\rho^4} } = \frac{1}{2 EL_0} \rho^{1/2}
\cosh^{3/2} \chi ~e^A \dot{r},
\]
\be
F(u) = e^A \sqrt{ \frac{\cosh \chi}{\rho} }, \qquad G(u) = \sqrt{
\frac{{\rm sech} \chi}{\rho^3} }, \qquad H(u) = \frac{1}{2} \sqrt{
\rho^3 ~{\rm sech} \chi},
\ee

\noindent and consider the metric
\[
\d s^2 = 2\d u \d v + E(u)^2 \d x^2 + F(u)^2 \d x^i \d x^i + G(u)^2 \d
z^1 \d \bar{z}^1 + H(u)^2 \d z^2 \d \bar{z}^2
\]
\be
\qquad \qquad \qquad \qquad \qquad \qquad - \frac{h^2}{4}{\cosh \chi} \left(\rho^3 |z^1|^2 + \frac{|z^2|^2}{4}
\right) \d u^2,
\label{eqn:gen_metric}
\ee

\noindent where $i=1,2,3$ and $z^1,z^2$ are complex coordinates on the
obvious $\mathbb{E}^2$s.  Then, with a dot
denoting $\del /\del u$, the relevant Brinkman coordinates are
\[
\hat{u} = u, \qquad \hat{x} = E x, \qquad \hat{x}^i = F x^i, \qquad \hat{z}^1 = G z^1, \qquad \hat{z}^2 = H z^2,
\]
\be
\hat{v} = v - \frac{1}{2} \left( E \dot{E} x^2 + F \dot{F} x^i x^i + G
\dot{G} |z^1|^2 + H \dot{H} |z^2|^2 \right),
\label{eqn:brinkman}
\ee

\noindent in terms of which the metric (\ref{eqn:gen_metric}) becomes,
dropping the hats,
\be
\d s^2 = 2\d u \d v + \d s^2(\mathbb{E}^8) - \left[
- \frac{\ddot{E}}{E} x^2 - \frac{\ddot{F}}{F} x^i x^i + \left(
\frac{h^2}{4} \rho^3 - \frac{\ddot{G}}{G} \right) |z^1|^2 + \left(
\frac{h^2}{16} \frac{1}{\rho^3} - \frac{\ddot{H}}{H} \right) |z^2|^2
  \right] \d u^2.
\ee

\noindent We will not consider the form fields explicitly, but it is
easy to see that an application of the Penrose limit will give the
same fields (\ref{wavefields}) as for the IR solution of section
\ref{sec:IR}, but with a $u$--dependent amplitude.

At any rate, the resulting metric is certainly in the form of a one--half supersymmetric
pp--wave, but with a complicated $u$--dependent profile.  It seems
unlikely that string theory on this background is tractable.
Moreover, it is somewhat difficult to see what statements about
the dual gauge theory can be made.  The immediate observation in this
regard is, of course, that there is no concept of operators with a
definite conformal dimension at a general point along the flow.
However, in the maximally supersymmetric case, dual to the $\N=4$
Yang--Mills theory, we know~\cite{KP,DGR,BN} that evolution in light--cone time $u$
corresponds to changes of scale in the gauge theory (the original
holographic radial direction is a monotonic function of $u$).  It is
thus tempting to argue that string theory on the above pp--wave is
dual to an ``RG flow'' between the Penrose limit of the $\N=4$
Yang--Mills theory ($u = \infty$) and the Penrose limit of the $\N=1$
fixed point theory ($u=-\infty$).  Evolution in light--cone time would
then induce a flow between the relevant sectors of the two gauge
theories.

However, the interpretation must be more subtle as is
apparent from considering the Penrose limit of
AdS$_5\times S^5$ in Poincar\'e coordinates. For example
in (\ref{eqn:r_and_t}), one finds
that the usual null trajectories start at $r=0$,
travel out to some maximum $r=L \ln(E/h)$ and then
fall back to $r=0$. Hence these geodesics sample a finite
range of energies extending from the far IR to some maximum,
which depends solely on the choice of the initial conditions
for the geodesic. Clearly, unravelling information about the
RG flow with the analogous geodesics above is a challenging
but potentially fruitful problem. We should also add that
similar geodesics in nonconformal backgrounds were considered
in refs.~\cite{ncon1,ncon2}, and a discussion of RG flows in
this context also appeared in ref.~\cite{DGR}.




\vspace{1cm}
\noindent
{\bf Acknowledgements}\\
DB and KJL would like to thank Bert Janssen, David Page, Simon Ross,
Paul Saffin and Douglas Smith for conversations.  CVJ would like to thank the
organisers of the 2002 KIAS workshop on Strings and Branes for
inviting him to present a seminar on this material, and also thank
Matthias Gaberdiel, Petr Ho\u{r}ava, Nakwoo Kim, Shiraz Minwalla and,
along with DB, David Tong for useful comments and discussions.  CVJ and RCM thank the Isaac
Newton Institute for Mathematical Sciences for their hospitality in the
initial stages of this project.  DB is supported in part by the EPSRC
grant GR/N34840/01, and KJL by an EPSRC studentship. CVJ thanks the
EPSRC for support.  Research by RCM is supported in part by NSERC of
Canada and Fonds FCAR du Qu\'ebec.


\appendix

\renewcommand{\theequation}{\Alph{section}.\arabic{equation}}

\sect{Appendix: the Pilch--Warner geometry}

The ten--dimensional Pilch--Warner geometry has the metric~\cite{PW1,PW2}
\be
\d s^2 = \Om^2 \left( e^{2A} \d s^2(\mathbb{M}^4) + \d r^2 \right)
+ \d s^2_5,
\ee

\noindent where, in terms of Cartesian coordinates $x^I, I=1,\ldots,6$,
on $\mathbb{E}^6$ such that $x^Ix^I=1$, the five-dimensional
``internal'' metric is
\be
\d s^2_5 = L_0^2 \frac{\mbox{sech} \chi}{\xi^3} \left[ \xi^2 \d x^I
Q_{IJ}^{-1} \d x^J + \sinh^2 \chi \left( x^I J_{IJ} \d x^J \right)^2 \right],
\ee

\noindent $L_0$ being the radius of the AdS space at the UV fixed
point.  The complex structure $J_{IJ}=-J_{JI}$ has non-zero components $J_{14} =
J_{23} = J_{65} = 1$ and
\be
\Om^2 = \xi \cosh \chi, \qquad \xi^2 = x^I Q_{IJ} x^J, \qquad Q = \mbox{diag} \left( \rho^{-2}, \rho^{-2},
\rho^{-2}, \rho^{-2},\rho^4,\rho^4 \right).
\ee

\noindent The supergravity scalars $\chi(r)$ and $\rho(r)$ obey,
together with the metric function $A(r)$, the following equations~\cite{FGPW}:
\ba
\frac{\d \rho}{\d r} &=& \frac{1}{6L_0\rho^2} \left( \cosh(2\chi) (\rho^6+1) -
(3\rho^6-1) \right), \nonumber  \\
\frac{\d \chi}{\d r} &=& \frac{1}{2L_0\rho^2} \sinh(2\chi) (\rho^6-2), \\
\frac{\d A}{\d r} &=& -\frac{1}{6L_0\rho^2} \left( \cosh(2\chi)(\rho^6-2) -
(3\rho^6+2) \right). \nonumber
\ea

\noindent As explained in ref.~\cite{PW1}, one uses the complex coordinates
\be
u^1 = x^1 + i x^4, \qquad u^2 = x^2 + i x^3, \qquad u^3 = x^5 - i x^6,
\ee

\noindent parametrised as
\be
\left( \begin{array}{c} u^1 \\ u^2 \end{array} \right) = e^{-i\p/2}
\cos \th ~g \left( \begin{array}{c} 1 \\ 0 \end{array} \right), \qquad
u^3 = e^{-i\p} \sin \th,
\label{eqn:u}
\ee

\noindent where
\be
g = \left( \begin{array}{cc} v^1 & -\bar{v}^2 \\ v^2 & \bar{v}^1
\end{array} \right) \in SU(2),
\label{eqn:g}
\ee

\noindent is used to define the $SU(2)$ left--invariant one--forms, $\s_i$, in terms
of the Pauli matrices, $\t_i$, via
\be
\s_i=-\frac{i}{2} \Tr \left(\t_i ~g\inv \d g \right).
\ee

\noindent Our choice of these one--forms differs from that
of refs.~\cite{PW1,PW2} by a factor of $1/2$, which gives rise to various
discrepancies between the form-fields given here and
those in refs.~\cite{PW1,PW2}.  Our left-invariant one--forms satisfy $\d \s_i =
\varepsilon_{ijk} \s_j \wedge \s_k$ so the metric on the unit 3-sphere
is $\d\Om_3^2 = \s_i \s_i$.  In terms of Euler angles
$\{\al,\beta,\ga\}$ on the three--sphere, we have
\be
v^1 = \cos (\al/2) ~e^{i(\beta+\ga)/2}, \qquad v^2 = \sin (\al/2)
~e^{i(\beta-\ga)/2},
\ee

\noindent so that
\ba
\s_1 &=& \frac{i}{2} \left( \sin \beta \d \alpha - \cos \beta \sin
\alpha \d \ga \right), \nonumber \\
\s_2 &=& -\frac{1}{2} \left( \cos \beta \d \alpha + \sin \beta \sin
\alpha \d \ga \right), \\
\s_3 &=& \frac{1}{2} \left( \d \beta + \cos \alpha \d \ga
\right). \nonumber
\ea

\noindent There is thus a natural $U(1)$ action $\beta \ra \beta +
\mbox{const.}$, under which the $SU(2)$
doublet in (\ref{eqn:u}) picks up an overall
phase.  This global $U(1)_{\beta}$ rotates $\s_1$ into
$\s_2$ and leaves $\s_3$ invariant.  The local version can be used to
choose different parametrisation of the five--sphere directions.  Thus, to go from the
coordinates used in ref.~\cite{PW1} to those used in ref.~\cite{PW2}, one shifts
$\beta \ra \beta + \p$, which removes the overall phase from the
$SU(2)$ doublet in (\ref{eqn:u}), and induces the shift $\s_3 \ra
\s_3 + \p/2$.  There is a further $U(1)$ action $\ga \ra \ga +
\mbox{const.}$ which will be of
interest to us.  This $U(1)_{\ga}$ is the diagonal subgroup of the obvious global
$SU(2)$.  Roughly speaking, the doublet corresponds to the two massless
chiral superfields, $\Phi_1$ and $\Phi_2$, in the gauge theory, and the singlet corresponds to
the massive superfield $\Phi_3$.  More precisely~\cite{JLP1,JLP2}, the moduli space of
a D3-brane probe corresponds to the former, $\th=0$, directions and
the latter, $\th=0$, directions are orthogonal to this moduli space.

Putting this all together, we have (\emph{cf.} ref.~\cite{PW1})
\ba
\d s^2 &=& \frac{X^{1/2} \cosh \chi}{\rho} \left( e^{2A} \d s^2(\mathbb{M}^4)
+ \d r^2 \right) + \d s^2_5, \\
\d s^2_5 &=& L_0^2 \frac{X^{1/2}\mbox{sech} \chi}{\rho^3} \left[ \d\th^2 +
\frac{\rho^6 \cos^2 \th}{X} \left( \s_1^2 + \s_2^2 \right) +
\frac{\rho^{12} \sin^2 (2\th)}{4 X^2} \left( \s_3 +
\frac{(2-\rho^6)}{2\rho^6} \d \p \right)^2 \right. \nonumber \\
&& \qquad \qquad \left. + \frac{\rho^6 \cosh^2
\chi}{16 X^2} ( 3- \cos (2 \th) )^2 \left( \d \p -
\frac{4 \cos^2 \th}{(3- \cos (2 \th))} ~\s_3
\right)^2 \right],
\ea

\noindent where
\be
X(r,\th) = \cos^2 \th + \rho^6 \sin^2 \th.
\ee

\noindent Note that the global isometry group of the metric is
$SU(2) \times U(1)_{\beta} \times U(1)_{\p}$, although only a
combination of the two $U(1)$s is preserved by the form-fields.  In the
IR ($r \ra -\infty$), we have $\chi \ra 2/\sqrt{3}$, $\rho \ra
2^{1/6}$ and $A(r) \ra r/L$, where $L = (3/2^{5/3}) L_0$.  The metric
becomes~\cite{PW1}
\ba
\d s^2 (\mbox{IR}) &=& \frac{2^{1/3}}{\sqrt{3}} \left( 3- \cos 2\th
\right)^{1/2} \left( e^{2r/L} \d s^2(\mathbb{M}^4) + \d r^2 \right) +
\d s^2_5 (\mbox{IR}), \nonumber\\
\d s^2_5 (\mbox{IR}) &=& \frac{\sqrt{3} L_0^2}{4} \left( 3- \cos 2\th
\right)^{1/2} \left[ \d\th^2 + \frac{4 \cos^2 \th}{(3- \cos 2\th)} (\s_1^2 + \s_2^2) +
\frac{4 \sin^2 2\th}{(3- \cos 2\th)^2} \s_3^2 \right. \nonumber \\
&& \left. \qquad \qquad \qquad \qquad \qquad \qquad \qquad + \frac{2}{3} \left( \d\p
- \frac{4 \cos^2 \th}{(3- \cos 2\th)} \s_3 \right)^2 \right].
\label{eqn:app_metric}
\ea

\noindent Concentrating for the time being on this fixed point
geometry, our self--dual five--form is
\be
F_5 (\mbox{IR}) = - \frac{2}{3} \frac{2^{2/3}}{L} e^{4r/L} \left( 1 + \star
\right) \ep(\mathbb{E}^4) \wedge \d r,
\label{eqn:app_5-form}
\ee

\noindent which differs by a factor of 2 to that in ref.~\cite{PW1}.  To
determine the correct Ansatz for the three--form, one
considers the linear $G_3 = \d u^1 \wedge \d u^2 \wedge \d u^3$, which
depends on $\p$ only through the overall phase $e^{-2i\p}$, and which
includes an overall factor of $(\s_1 + i \s_2 )$ (this is $(\s_1 - i
\s_2 )$ in ref.~\cite{PW1,PW2} due to different conventions for the
left--invariant one--forms).  The two--form potential is thus
\be
A_2 (\mbox{IR}) = A ~e^{-2i\p}~ \frac{L_0^2 \cos \th}{2} \left( \d \th -
\frac{2i\sin (2\th)}{(3-\cos (2\th))} \s_3 \right) \wedge (\s_1 + i
\s_2 ),
\ee

\noindent where there is an overall arbitrary constant phase, $A$
which is set to $-i$ in ref.~\cite{PW1}.  The field strength $G_3 = \d A_2$ is
\be
G_3 (\mbox{IR}) = iA ~e^{-2i\p}~ L_0^2 \cos \th \left( \d \th \wedge \d \p - \frac{8
\cos ^2 \th}{(3-\cos(2\th))^2} \d \th \wedge \s_3 - \frac{2i \sin
(2\th)}{(3-\cos (2\th))} \s_3 \wedge \d \p
\right) \wedge \left( \s_1 + i \s_2 \right).
\label{eqn:app_3-form}
\ee

\noindent It should be obvious that the global $U(1)$ symmetry group of the
solution as a whole is the combination $U(1)_R = U(1)_{\p} + 2
U(1)_{\beta}$.  Of course, by shifting $\beta$ as discussed above,
one is free to choose the R-symmetry to be any combination of the two
$U(1)$s.  For example, in the text we are interested in
coordinates for which $U(1)_R = U(1)_{\p}$, so we perform the
coordinate transformation $\beta \ra \beta + 2 \p$ on the above
solution.  This removes the overall $\p$--dependent phase in the two--form
potential precisely as required.


\sect{Appendix: the $\th=\pi/2$ geodesics}

Here we will consider the Penrose limit of the IR fixed point solution along a null geodesic with
$\th=\pi/2$, corresponding to the massive direction orthogonal to the
moduli space.  We start with the solution (\ref{eqn:PW1}--\ref{eqn:PW3}), and take
\be
\rho = \frac{r}{L}, \qquad \th = \frac{\pi}{2} + \frac{y}{L},
\ee

\noindent with $L \ra \infty$.  Defining the light--cone coordinates,
\be
u = \frac{1}{2E} \left( \t + \frac{2}{3} \p \right), \qquad v =
-\frac{2^{4/3}}{\sqrt{3}} E L^2 \left( \t - \frac{2}{3} \p \right).
\ee

\noindent the solution becomes
\ba
\d s^2_2 &=& 2 \d u \d v + \frac{E^2}{4} \left( 1 - 4(r^2 + |v|^2)
\right) \d u^2 + \d s^2(\mathbb{E}^8) + iE ~\d u~(v^1 \d \bar{v}^1 +
v^2 \d \bar{v}^2), \nonumber \\
F_5 &=& - \frac{E}{2} \d u \wedge \ep( \mathbb{E}^4 ), \\
G_3 &=& - \sqrt{3} E~ \d u \wedge \d v^1 \wedge \d v^2. \nonumber
\ea

\noindent where we have rescaled $r$ and $y$ and defined the coordinates
\be
v^1 = e^{-iEu/2} z^1, \qquad v^2 = e^{-iEu/2} z^2,
\ee

\noindent $z^1,z^2$ being complex coordinates on the two
$\mathbb{E}^4$s.  The constant in $g_{uu}$ is unimportant as far as the field
equations and supersymmetry transformations are concerned, and has
been discussed in ref.~\cite{CLP1}.  The metric has a $\d u \d x^i$
cross--term, but it is clear that this can be traded with explicit
$u$--dependence in the three--form.

In the original coordinates, that is, the cross term is of the form $\s_3~ \d u$ and
can be removed by shifting the Euler
angle $\beta$, to give
\ba
\d s^2_2 &=& 2 \d u \d v - E^2 (r^2 + |z|^2) \d u^2 + \d
s^2(\mathbb{E}^8), \nonumber \\
F_5 &=& - \frac{E}{2} \d u \wedge \ep( \mathbb{E}^4 ), \\
G_3 &=& - \sqrt{3} E e^{-iEu} \d u \wedge \d z^1 \wedge \d z^2. \nonumber
\ea

\noindent Substituting for
$\la = -E/2$ and $\mu = -\sqrt{3} E
~e^{-iEu}$ in the field equation (\ref{eqn:trace_A}), one can verify
that the solution is valid.  The $u$--dependence in $G_3$ drops out
of the field equations since it is just an overall phase.  We note
that the above pp--wave will give rise to worldsheet scalars of the
same mass, unlike the case we have considered in the text.

We have not been able to understand the significance of this
particular Penrose limit with respect to the dual gauge theory.  In
the original coordinates with a cross term in the metric, the
light--cone Hamiltonian one finds is
\be
H = \Delta - \frac{3}{2} J,
\ee

\noindent so that all six scalar fields have $H=0$ --- there is
certainly no unique ground state.  Moreover, this fact does not seem
to be mirrored in the string theory spectrum on this background, which
\emph{does} seem to show a unique ground state.  Furthermore, the
frequencies of the bosonic and fermionic modes do not seem to match,
in which case it seems unlikely that a simple Hamiltonian can be
written down at all.  On the other hand, after shifting $\beta$ to remove the
cross--term, one finds an $i \del_{\beta}$ term in the Hamiltonian
and, as discussed in section \ref{sec:ft}, there is no conserved charge
in the gauge theory associated with this differential operator.

Taking the Penrose limit along geodesics with angular
momentum in the massive directions is perhaps an odd thing to try to
do anyway, since at the IR fixed point, one can simply integrate out these
directions.  As far as a D-brane probe would be concerned, motion in these
directions is energetically disfavoured and simply not to be described in the dual picture
by the effective low energy $\N=1$ field theory.


\newcommand{\gsim}{\mathrel{\raisebox{-.6ex}{$\stackrel{\textstyle>}{\sim}$}}}
\newcommand{\lsim}{\mathrel{\raisebox{-.6ex}{$\stackrel{\textstyle<}{\sim}$}}}

\section{Appendix: Instabilities for large $B$--fields}

Let us consider the general solution (\ref{eqn:ansatz}), with a
diagonal matrix $A_{ij} = -\delta_{ij} E_i^2$, for which the supergravity
equations of motion (\ref{eqn:trace_A}) yield:
\be
\sum E_i^2=8\mu^2+2|\zeta|^2.
\label{eqn:neutrace}
\ee
Following the analysis of section 4.1, the equations of motion for $X^1$
and $X^3$ are then
\ba
\nabla^2 X^1 - M_1^2 X^1 + b \del_\s X^3 & = & 0 , \\
\nabla^2 X^3 - M_3^2 X^3 - b \del_\s X^1 & = & 0 ,
\ea
where $b = p^+ \al' \zeta$ and $M_i = p^+ \al' E_i$.
(Similar equations hold for the $X^2$ and $X^4$ directions.)
Fourier expanding as in (\ref{eqn:fourier}) above, one finds that
the frequencies of the normal modes are:
\be
\omega_n^2 = n^2 + \frac{M_1^2 + M_3^2}{2}
\pm\sqrt{{1\over4}(M_1^2 - M_3^2)^2+b^2n^2}
\label{eqn:allthatjazz}
\ee
Now note that \emph{all} of the above $\omega_n$'s are real and
non--zero if and only if
\be b^2 < n^2 + M_1^2 + M_3^2 + \frac{M_1^2M_3^2}{n^2}. \ee
In particular, by minimising the right-hand side
with respect to $n$, one is guaranteed real $\omega_n$ for
\be b^2< (M_1 + M_3)^2.
\label{eqn:bigB}
\ee
However, for larger values of $b^2$, it \emph{is} possible that some
of the frequencies are imaginary, resulting in exponentially
growing string modes.

Generalising the analysis of section 4.2, one finds no such
instability in the fermionic spectrum. As well as the standard four
fermions with frequency $\omega^2_n = n^2 + M^2$ at each level, the
remaining fermionic oscillators have \be \omega_n^2 = n^2 + M^2 +
\frac{b^2}{2} \pm\frac{1}{2}\sqrt{b^4+4b^2n^2} \ee where $M = p^+
\al' \mu$. It is straightforward to show that this expression always
yields real frequencies.  Note that the metric coefficients $E_i^2$ do
not appear directly in the fermionic spectrum. Further, for the
general background, the bosonic and fermionic spectra no longer match.

The interpretation of the unstable modes is somewhat unclear, although
their existence is quite interesting.  One might think
of them as some sort of (classical) instability in the string
theory in these backgrounds. The appearance of these modes is
particularly curious because the supergravity background still
appears to be at least one--half supersymmetric, \emph{i.e.}, the 16 standard
Killing spinors annihilated by $\Gamma_+$ will yield vanishing
supersymmetry variations (\ref{eqn:delta_fermions}), irrespective of
the value of the three--form field. One might suspect that these
Killing spinors are ill--behaved in some way, {\it e.g.}, exponentially
diverging in $u$, if the three--form is too large. However, this
is not the case as is obvious since as noted above the equation
for standard Killing spinors is independent of $G_3$.

One might imagine that solutions with the $B$--field too large, in the
above sense, are excluded by the supergravity field equations
(\ref{eqn:neutrace}).  However, it is easy to see that this is not the
case, as the inequality (\ref{eqn:bigB}) only refers to three of the
nine parameters appearing in the former equation. Hence these unstable
modes apparently appear in valid supergravity backgrounds. Further it
seems that given the null form of these gravity wave solutions, they
will be solutions of the string equations of motion to all
orders in $\alpha'$ \cite{Amati,horror1,horror2}. In particular, it
seems the general discussion of ref.~\cite{horror1} should apply even with
the appearance of R--R fields in the background.

Let us make several further observations. First, these solutions are
in no sense asymptotically flat in any directions, rather the field
strengths and $R_{uiuj}$ are constant throughout the spacetime. Hence
one may wonder whether or not the generic backgrounds are relevant in
string theory. Certainly we have found that certain pp--wave solutions
(in fact a very broad family, given the results of section 6 and
appendix B) appear as the Penrose limits in asymptotically AdS
backgrounds, and so play a role in string theory. It could be that
these ``unstable'' supergravity solutions are simply pathological
backgrounds as far as the string theory is concerned and are not
useful spacetimes to consider from this point of view. A second
observation is that the unstable modes in the bosonic spectrum only
occur at finite, non--zero $n$. Roughly, one may think that oscillator modes need
to be excited so that the string is spatially extended and can ``see''
the $B$--field. Hence the instability is inherently stringy in origin.
This feature is somewhat reminiscent of the instabilities discussed in
ref.~\cite{cow}.  Finally, we note that the instability only appears for a
finite set of modes, \emph{i.e.}, for a finite range of $n^2$. It is
straightforward to derive the exact range, however, let us make some
qualitative statements. Generically if we take $M_1^2\simeq
M_3^2\simeq M^2$ then the instability sets in for $b^2\gsim M^2$. In
this case the unstable modes appear in a certain range,
$n_-^2<n^2<n_+^2$, where $n^2_\pm=\O(M^2)$ and $n^2_+-n_-^2=\O(M^2)$.
However, recall the definitions above, $M_i^2=(p^+ \al' E_i)^2$. Now
in studying supergravity backgrounds, we would ask that typical
curvatures are small which in this case corresponds to $(l_s E_i)^2\ll
1$. If this inequality applies and $(l_sp^+)^2\lsim 1$, then the
unstable range will lie entirely within the range 0 and 1. That is,
there will not actually be any integer values of $n$ for which the
frequencies (\ref{eqn:allthatjazz}) become imaginary. Hence the
appearance of an actual unstable mode requires that either the
background is highly curved on the string scale and/or the $p^+$
component of the momentum is very large (which corresponds to a highly
excited string state).  
This once again emphasises the stringy nature of this potential instability.


\sect{Appendix: Penrose limit of AdS$_5 \times S^5$ in Poincar\'{e} coordinates}

Although the Penrose limit of AdS$_5$ $ \times S^5$ in Poincar\'{e}
coordinates has already been discussed in ref.~\cite{BFP}, it is worth
reviewing the analysis here: firstly, we use different coordinates on
the five--sphere, which leads initially to a ``mixed'' Rosen--Brinkman form
of the maximally supersymmetric pp--wave; and secondly, it will be
useful to compare this simple case with the more complicated geometry
to follow.  The metric on AdS$_5$ $\times S^5$, with the $AdS$ factor in
global coordinates, is
\be
\d s^2 = L^2 \left[ -\cosh^2 \rho ~\d\t^2 + \d\rho^2 + \sinh^2 \rho ~\d\Om^2_3 +
\cos^2 \th ~\d\p^2 + \d\th^2 + \sin^2 \th ~\d\hat{\Om}^2_3 \right],
\ee

\noindent where $\d\Om^2_3$ and $\d\hat{\Om}^2_3$ denote metrics on
a unit three--sphere.  In these coordinates, a simple class of null
geodesics is that for which $\rho = 0 =\th$.  Taking the Penrose
limit along such a geodesic which has angular momentum in the $\p$
direction gives rise~\cite{BFHP2,BFP,BMN} to the maximally supersymmetric
pp--wave of type IIB supergravity~\cite{BFHP1}.  Poincar\'{e} coordinates on
AdS$_5$ are defined by
\ba
y&=&\frac{1}{L}(\cosh \rho ~\cos \t - \sinh \rho ~\Om_4),
\label{eqn:y} \nonumber \\
t&=&\frac{1}{y}\cosh \rho ~\sin \t, \\
x^i&=&\frac{1}{y} \sinh \rho ~\Om_i, \nonumber
\ea

\noindent where $x^i, i=1,2,3$, are the coordinates on
$\mathbb{E}^3$ and where $\Om_i \Om_i + \Om_4 \Om_4 =
1$ gives an embedding of $S^3$ in $\mathbb{E}^4$.  Defining a new
radial coordinate
\be
r=L \ln (Ly),
\ee

\noindent the metric on AdS$_5 \times S^5$ is thus
\be
\d s^2 = e^{2r/L} \left[-\d t^2 + \d s^2(\mathbb{E}^3) \right] + \d r^2 + L^2 \left[
\cos^2 \th ~\d\p^2 + \d\th^2 + \sin^2 \th ~\d\hat{\Om}^2_3 \right].
\label{eqn:metric}
\ee

For geodesics at a constant point in $\mathbb{E}^3$, the effective
Lagrangian is
\be
\L = - e^{2r/L} \dot{t}^2 + \dot{r}^2 + L^2 \left[ \cos^2 \th
~\dot{\p}^2 + \dot{\th}^2 + \sin^2 \th \dot{\hat{\Om}}^2_3 \right],
\ee

\noindent where, if $\la$ is the affine parameter,  a dot denotes
$\d/\d\la$.  One is still free to consider the class of geodesics for
which $\th=0$.  Then the $t$ and $\p$ equations give
\be
\dot{t} = ELe^{-2r/L}, \qquad \dot{\p}=h,
\label{eqn:tdot}
\ee

\noindent where $E$ and $h$ are the conserved energy and angular
momentum associated with the Killing vectors $\del/\del t$ and $\del
/\del \p$ respectively.  The null condition $\L=0$ then gives
\be
\dot{r} = \pm EL \sqrt{e^{-2r/L} - \frac{h^2}{E^2}}.
\label{eqn:rdot}
\ee

\noindent If we choose the $-$ sign in the above, then the
resulting geodesic matches onto the $\rho = 0$ geodesics in global
coordinates.\footnote{The plus sign gives rise to a second class of
null geodesics, the Penrose limit along which has precisely the same
effect on the spacetime, so we will not consider it here.}
Integrating (\ref{eqn:rdot}) and (\ref{eqn:tdot}) gives
\be
r(\la) = L \ln \left( \frac{E}{h} \cos \la h \right), \qquad t(\la) =
L \frac{h}{E} \tan \la h.
\label{eqn:r_and_t}
\ee

\noindent Transforming back to the $y$ coordinate defined in
(\ref{eqn:y}) above, gives
\be
y(\la) = \frac{1}{L} \frac{E}{h} \cos \la h,
\ee

\noindent which, with $h=E$, matches onto the $\rho=0$ null
geodesics in global coordinates, as promised (and these latter do
indeed have $h=E$).

Following ref.~\cite{BFP}, we introduce coordinates $\{u,v,x\}$ such that $u$
is the affine parameter along the null geodesics.  Demanding that
$g_{uu} = 0 = g_{ux}$ and $g_{uv}=1$, a possible choice is
\ba
\del_u &=& \dot{r} \del_r + \dot{t} \del_t + \dot{\p} \del_{\p},
\nonumber\\
&=& -Lh \tan u h ~\del_r + \frac{h^2L}{E} \sec^2 u h ~\del_t + h
\del_{\p}, \label{eqn:coords1} \nonumber \\
\del_v &=& -\frac{1}{EL} \del_t, \\
\del_x &=& \frac{1}{L} \del_{\p} + \frac{h}{E} \del_t, \nonumber
\label{eqn:coords2}
\ea

\noindent which can be integrated to give
\ba
r (u) &=& L \ln \left( \frac{E}{h} \cos uh \right), \nonumber \\
t (u,v,x) &=& \frac{hL}{E} \tan uh - \frac{v}{EL} + \frac{h}{E} x, \\
\p (u,x) &=& \frac{x}{L} + hu. \nonumber
\ea

We now write the original metric (\ref{eqn:metric}) in terms of
$\{u,v,x\}$ and implement the fact that $\th=0$ by defining $\th =
y/L$ and taking the limit $L \ra \infty$.  Dropping terms of $\O
(1/L)$ gives the metric
\be
\d s^2 = 2\d u \d v - h^2 y^2 \d u^2 + \sin^2 uh ~\d x^2 + \frac{E^2}{h^2} \cos^2 uh
~\d s^2(\mathbb{E}^3) + \d s^2(\mathbb{E}^4),
\label{eqn:pp_mixed}
\ee

\noindent where $y$ is the radial coordinate on $\mathbb{E}^4$.
The coordinate singularities in this metric appear
because of degeneracies in the choice of vectors in
(C.10). For example, $\partial_u=hL\partial_x$ at
$\sin uh=0$. Further, we
note in passing that working in global coordinates, and the analogue
thereof on the sphere, gives rise to the pp--wave in Brinkman
coordinates.  Use of Poincar\'{e} coordinates, however, and the
parametrisation of the five--sphere used in ref.~\cite{BFP}, gives rise to the
pp--wave in Rosen coordinates.  The ``mixed'' coordinates used here has
given rise to the above pp--wave in ``mixed'' Brinkman--Rosen
coordinates.  At any rate, introducing
\[
x^- = u, \qquad z = \sin uh ~x, \qquad  z^i = \frac{E}{h} \cos uh
~x^i, \nonumber
\]
\be
x^+ = v + \frac{1}{4} \left( \frac{E}{h} x^i x^i -x^2 \right) \sin (2uh),
\ee

\noindent gives
\be
\d s^2 = 2 \d x^- \d x^+ - h^2 |x|^2 \d x^{-2} + \d s^2 (\mathbb{E}^8),
\label{eqn:pp_brink}
\ee

\noindent where now $|x|$ denotes the radial coordinate on $\mathbb{E}^8$.

As to the R--R five--form field strength which, in Poincar\'{e} coordinates,
has the form
\be
F_5 = \frac{C}{L} (1 + \star) ~\ep (AdS_5) = \frac{C}{L} e^{4r/L} (1+
\star) ~\d t \wedge \ep(\mathbb{E}^3) \wedge \d r,
\ee

\noindent for some constant $C$ and where $\ep(\M)$ denotes the volume
form on $\M$, we find
\be
F_5 = C \d x^- \wedge \ep (\mathbb{E}^4).
\label{eqn:5-form}
\ee



\end{document}